\documentclass[a4paper,11pt]{article}
\usepackage{jcappub} 
\usepackage{graphicx,subfigure}
\usepackage{amsmath}
\usepackage{placeins}
\usepackage{amssymb}
\usepackage{longtable}
\usepackage{multirow}
\usepackage[flushleft]{threeparttable}
\usepackage{booktabs}
\usepackage{lineno}
\usepackage{chngpage}
\usepackage{geometry}
\usepackage{epstopdf}

\title{Search for correlations between the arrival directions of IceCube neutrino events and ultrahigh-energy cosmic rays detected by the Pierre Auger Observatory and the Telescope Array}
\collaboration{The IceCube, Pierre Auger and Telescope Array Collaborations}

\abstract{This paper presents the results of different searches for correlations between very high-energy neutrino candidates detected by IceCube and the highest-energy cosmic rays measured by the Pierre Auger Observatory and the Telescope Array. We first consider samples of cascade neutrino events and of high-energy neutrino-induced muon tracks, which provided evidence for a neutrino flux of astrophysical origin, and study their cross-correlation with the ultrahigh-energy cosmic ray (UHECR) samples as a function of angular separation. We also study their possible directional correlations using a likelihood method stacking the neutrino arrival directions and adopting different assumptions on the size of the UHECR magnetic deflections. Finally, we perform another likelihood analysis stacking the UHECR directions and using a sample of through-going muon tracks optimized for neutrino point-source searches with sub-degree angular resolution. No indications of correlations at discovery level are obtained for any of the searches performed. The smallest of the p-values comes from the search for correlation between UHECRs with IceCube high-energy cascades, a result that should continue to be monitored.}

\keywords{neutrino experiments, ultra high energy cosmic rays, cosmic ray experiments, neutrino astronomy}

\begin{document}
\maketitle
\flushbottom

%\linenumbers
\section{Introduction}

The origin of the ultrahigh-energy cosmic rays is one of the main open questions in high-energy astrophysics. It is likely that they are accelerated in astrophysical objects, since more exotic scenarios for their origin, such as the top-down models, are strongly constrained by stringent limits on the flux of primary photons at such energies \cite{augerphotons,taphotons}. However, their sources still remain undiscovered after five decades of experimental efforts that have culminated in the two largest UHECR experiments ever built, the Pierre Auger Observatory \cite{augernim04,augernim15} and the Telescope Array \cite{AbuZayyad:2012kk}, in the Southern and Northern hemispheres, respectively. The two experiments, which measure UHECRs through the detection of extensive air showers produced in the Earth's atmosphere, have collected in total more than 300 events with energies in excess of $\sim 50$~EeV (where 1~${\rm EeV}\equiv 10^{18}$~eV). These energies are well within the region of very significant cosmic-ray flux suppression observed by both experiments \cite{augerspectrum,taspectrum}. If this suppression were due to the degradation of the cosmic-ray (CR) energies by their interactions with the cosmic radiation backgrounds \cite{gzk}, one would expect that the sources of such energetic CRs are relatively nearby, at distances shorter than $\sim 200$~Mpc. However, the results of several studies of UHECR arrival direction distributions \cite{augerapj2015,taardir} have shown no significant small-scale anisotropy and have not allowed the establishment of statistically-significant correlations with nearby astrophysical sources. 

The identification of UHECR sources by correlation analyses is a challenging task. Since UHECRs are mostly charged particles, their trajectories from their sources to the Earth are deviated by magnetic fields, most notably by the Galactic field. The deviations depend on the energy and charge of the cosmic ray: while at energies above 50 EeV they may be of order of a few degrees in the case of protons, such deviations would be much larger for nuclei of the same energy but having a larger charge. This situation is further complicated by the facts that there are still large uncertainties in the modeling of magnetic fields (see \cite{farrarcc2014,Haverkorn:2014jka} and references therein) and that the UHECR mass composition at energies above 50 EeV is still largely unexplored \cite{augerxmax,taxmax}. 

Further, UHECRs accelerated in astrophysical sources are naturally expected to produce high-energy photons and neutrinos in interactions with the ambient matter and radiation. In contrast to UHECRs, photons and neutrinos are not subject to magnetic deflections in their propagation to the Earth and hence they point directly to their sources. However, high-energy photons may also be produced in purely leptonic processes, and hence may be unrelated to hadronic CRs. Moreover, photons at energies exceeding a few TeV can be significantly attenuated by $e^+e^-$-pair production in interactions with the Cosmic Microwave Background (CMB) and other radiation backgrounds as they propagate from faraway sources. On the other hand, neutrinos interact with matter very weakly, so that they travel essentially unattenuated, and hence are excellent tracers of the sources of cosmic rays. Due to the challenge of neutrino detection, neutrino astronomy has however become viable only very recently, thanks to the discovery by the IceCube neutrino telescope \cite{HESE2paper} of a flux of extraterrestrial high-energy neutrinos extending up to at least 2~PeV.

The observations of the highest-energy events, on the one hand in neutrinos by IceCube and on the other, UHECRs, by the Auger Observatory and the Telescope Array, set the stage for a joint analysis of the three Collaborations presented in this paper, namely the search for a possible association between the IceCube neutrinos and the UHECRs. Some analyses searching for possible correlations have been presented in \cite{olinto, razzaque,gago}, but using smaller data sets.

Attempting to exploit the IceCube neutrino observations to identify the sources of UHECRs is motivated by several reasons. First, the arrival directions of the IceCube high-energy sample, in which a cosmic component has been identified \cite{HESE2paper}, is consistent with an isotropic distribution and therefore with the hypothesis of an extragalactic origin for the bulk of the events, although the presence of a Galactic component is possible. Second, the inferred diffuse flux of astrophysical neutrinos, with energies between 30~TeV and 2~PeV, is at the level of the Waxman-Bahcall flux \cite{waxbah}. This is the upper limit for the extragalactic neutrino flux calculated assuming that UHECRs are protons escaping from optically thin sources with an injection spectrum scaling as $E^{-2}$ and normalized to the measured UHECR flux above EeV energies. Clearly, if the sources are not optically-thin, or if the injection spectrum differs from $E^{-2}$, or the UHECRs are not protons, the upper limit on the flux would be considerably modified \cite{mannheim}. 

If neutrinos result from the decays of pions produced in $p\gamma$ or $pp$ processes, they would carry about 3--5\% of the proton energy. Hence, the neutrinos observed by IceCube with energies of 30 TeV to 2 PeV would have been produced in this case by protons with PeV--100 PeV energies. Although these energies are much smaller than those of the UHECRs considered here, the third argument for our analysis is based on the fact that it is only at the highest energies that CR trajectories could point to the sources, while at much lower energies CRs diffuse and may not even arrive to the Earth from faraway sources. Since UHECR sources will also produce lower energy CRs, it is possible that they are the same sources that produced the observed neutrinos. One should keep in mind, however, that there may also be sources able to produce the PeV neutrinos but which do not accelerate CRs up to ultrahigh-energies. We also note that electron antineutrinos produced in the $\beta$ decay of neutrons (or of radioactive nuclei produced in photo-disintegration processes) would carry about $4\times 10^{-4}$ of the energy of the parent nucleon, and hence PeV neutrinos could in principle also arise from EeV neutrons.

In this paper we present several searches for correlations between the arrival directions of UHECRs detected by the Auger Observatory and by the Telescope Array, and of different sets of events observed by the IceCube Neutrino Observatory. The main characteristics of the three detectors, and their respective data sets, are described in Section 2. The Auger and Telescope Array samples include 231 events with $E\geq 52$ EeV and 87 with $E\geq 57$ EeV, respectively. The IceCube data sets that we consider, obtained using different selection criteria, comprise 39 cascades (signatures of charged-current $\nu_e$ interactions as well as neutral-current interactions of all flavors) and 7 high-energy tracks (signatures of charged-current $\nu_\mu$ interactions) of the so-called `High Energy Starting Events' (HESE) \cite{HESE2paper,HESE3paper,HESE4yrICRC}, as well as 9 high-energy muon tracks whose extraterrestrial origin is highly probable \cite{weaver}. We also exploit the so-called point-source sample of about 400,000 tracks with a sub-degree angular resolution \cite{IC86PSPaper}.

The search for correlations between neutrinos and UHECRs is not straightforward. On the one hand, as noted above, due to the magnetic deviations discussed in Section~\ref{sec:magdef}, the UHECR arrival directions do not exactly correspond to the source positions. On the other hand, the neutrino HESE cascade events are characterized by a large angular uncertainty. Moreover, the IceCube samples are not exempt from backgrounds due to atmospheric secondary particles. In particular, the point-source sample is largely dominated by atmospheric muons in the Southern hemisphere and by atmospheric neutrinos in the Northern one. To account for these aspects, three different analyses have been devised.  Two analyses use the HESE and high-energy muon samples (Section~\ref{sec:UHECR-HESE}). The first one is a cross-correlation analysis where the number of pairs between CRs and the high-energy neutrino sample is evaluated within a large range of angular windows, from $1^\circ$ up to $30^\circ$. The second one adopts a stacking likelihood analysis, summing the contributions from the different sources. In this case, it is the arrival directions of the neutrinos that are being stacked. Plausible UHECR magnetic deflections and the neutrino point-spread functions are accounted for, together with the possible backgrounds. A third analysis uses a likelihood method applied to the point-source sample (Section~\ref{sec:pssample}). We note that the details of the analyses (angular scales explored, data samples considered, null hypotheses) were selected, as discussed in Sections 4.1, 4.2 and 5.1, before unblinding the analyses (see also \cite{Utah}), while some a posteriori cross checks are also mentioned. Finally, we conclude in Section~\ref{sec:concl}.

\section{The observatories and data sets}
\label{sec:datasets}
\subsection{The IceCube Neutrino Observatory}
The IceCube Observatory is a km$^3$-sized Cherenkov detector embedded in the ice at the geographic South Pole \cite{FirstYearPerformancePaper}. Optimized to detect neutrinos above $\sim 100$ GeV energies, it consists of 5160 photomultiplier tubes (PMTs) along 86 cables (called `strings')  instrumented between depths of 1450 and  2450\,m beneath the surface of the ice sheet. Each PMT is housed in a digital optical module (DOM), consisting of a pressure-resistant sphere with on-board digitization and Light Emitting Diodes (LEDs) for calibration~\cite{PMTPaper}. The DOMs detect Cherenkov photons emitted by charged leptons that traverse the ice volume~\cite{DOMMBPaper}. Starting in the year 2005, the detector ran in partial configurations consisting of 9, 22, 40, 59 and 79 strings until  its completion in December 2010. 

\subsubsection{High-energy starting events (HESE)}
\label{sec:HESE_sample}
In 2013, IceCube reported the first evidence for a high-energy neutrino flux of extraterrestrial origin from a search carried out on data collected between May 2010 and May 2012 in the 79-string and full 86-string configurations of the detector~\cite{HESE2paper}. That search, targeting high-energy neutrinos interacting within the IceCube detector, was later updated to include two more years of data taken with the full 86-string configuration, extending the observation period to May 2014~\cite{HESE3paper,HESE4yrICRC}. 
These searches have consistently demonstrated the existence of an astrophysical flux of neutrinos at the level of $E_{\nu}^2\phi(E_\nu ) \sim 10^{-8}\,{\rm GeV}\,{\rm cm}^{-2}\,{\rm s}^{-1}\,{\rm sr^{-1}}$ per flavor, emerging above atmospheric neutrinos starting at around 100\,TeV and extending to the PeV range. The hypothesis of a purely atmospheric explanation of the neutrino events collected during 4 years has been rejected at around $6.5\sigma$. However, the 54 events consisting of 39 cascades and 15 track-like events have shown no significant directional clustering, leaving unanswered the question of their origin~\cite{HESE3paper,HESE4yrICRC}. 

The median angular resolution of the events containing muon tracks is around~$1^\circ$, while the resolution of cascade-like events is around $15^\circ$ \cite{HESE2paper}. The expected atmospheric background for the HESE sample is predominantly track-like atmospheric muons not vetoed by external layers of DOMs and sneaking in through the region called the `dust layer'  (where dust obscures the transparency of ice) and from atmospheric $\nu_\mu$ charged-current interactions from the Northern hemisphere. We have therefore not used all the detected track-like events, but only the 7 that are more likely to be of extraterrestrial origin due to their high energies and/or arrival directions. Atmospheric neutrinos from the Southern hemisphere are accompanied by muons from the same shower and this background is greatly reduced by the event selection using a muon veto~\cite{HESE3paper}. 

Within this paper these 39 cascade events and 7 track-like events are referred to as `high-energy cascades' and `HESE tracks', respectively, and their properties are listed in Table~\ref{cascades_list} and Table~\ref{tracks_list} in the Appendix.

\subsubsection{High-energy through-going muons}
\label{sec:HETM_sample}
A data sample of~35,000 through-going muons from the Northern sky was selected from events recorded between May 2010 and May 2012 in a search for a diffuse up-going $\nu_\mu$ flux \cite{weaver}. While the majority of these events originate from neutrinos produced by cosmic-ray interactions in the Earth's atmosphere, at the highest energies a clear excess was observed. The hypothesis that the excess is of atmospheric origin was rejected at the 3.7$\sigma$ confidence level \cite{weaver}. This excess is compatible with the sum of the predicted muon neutrinos of atmospheric origin and an astrophysical flux consistent with the flux resulting from the HESE analysis. This deviation from the expectation that all events are of atmospheric origin occurs at the highest energies, where 9 events have been detected. These events (which are available in \cite{CWdatarelease}) are listed in Table~\ref{tracks_list} and have a median angular resolution of around $1^\circ$. The combination of these events with the 7 selected HESE tracks is referred to as `high-energy tracks'.

\subsubsection{4-year point-source sample}
\label{4pssamle}

High-energy muons have long tracks within IceCube and they can be reconstructed with a median angular resolution smaller than $1^{\circ}$, and hence they have good sensitivity for point-source searches. The corresponding effective volume to neutrino-induced muon events can be larger than the instrumented volume since the interaction vertex of the neutrino does not have to be contained within the detector. Four dedicated samples of such events have been isolated in three years of operation of IceCube in the incomplete 40-, 59- and 79-string configurations~\cite{IC79Paper, IC40Paper} as well as in one year of operation of the completed 86-string detector~\cite{IC86PSPaper}. Note that 7 events out of the 9 high-energy through-going muons described in Section~2.1.2 are contained in the point-source data set.

\begin{figure}[t]
  \centering
  \includegraphics[width=4.0in]{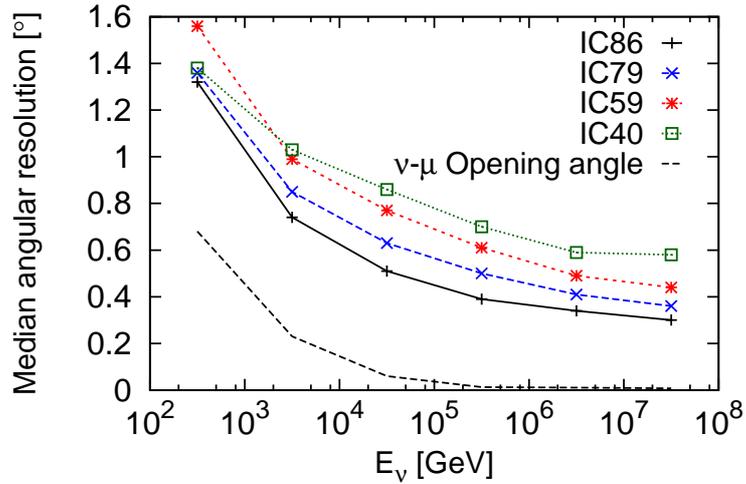}
  \caption{Muon neutrino angular resolution for the Northern sky (up-going events), defined as the median of the angle between the true neutrino direction and the reconstructed muon direction. The resolution is given for different analyzed samples as a function of the neutrino energy after all cuts used to select the final samples for the neutrino point-source searches. The kinematic angle between the muon and the neutrino is also shown with a dashed line.}
  \label{fig:SampledetailsPSF}
\end{figure}

\begin{figure}[h!]
  \centering
  \includegraphics[width=4.0in]{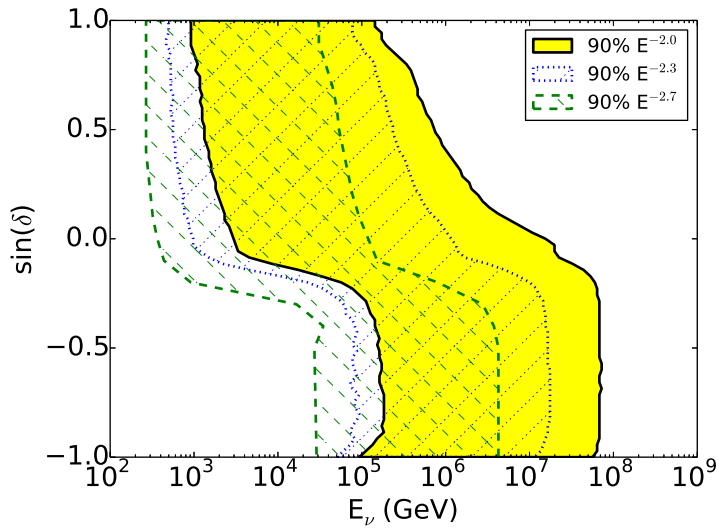}
  \caption{Neutrino energy interval containing 10\% to 90\% of the simulated signal events for different spectra and declinations for the four-year combined point-source samples after all analysis cuts are applied.}
  \label{fig:DecVsEnergy}
\end{figure}

The angular resolution of this sample has improved over the years, benefiting from the larger instrumented volumes achieved as the number of strings was increasing as well as from improvements in the reconstruction methods (see Figure~\ref{fig:SampledetailsPSF}). These samples are dominated by muons from charged-current interactions of atmospheric $\nu_{\mu}$ (and $\bar{\nu}_{\mu}$) in the Northern hemisphere (up-going data) and muons produced by interactions of cosmic rays with the Earth's atmosphere in the Southern hemisphere (down-going data). In the Northern sky, the effective area of the detector, determined by the analysis cuts and the opacity of the Earth for neutrinos with energies  above $\sim 100$~TeV, leads to a sample of events peaked in the 1\,TeV -- 1\,PeV range. In the Southern sky the sensitivity is in the 100\,TeV -- 100\,PeV range (see Figure~\ref{fig:DecVsEnergy}) due to the overwhelming background of atmospheric muons that needs to be reduced by imposing a higher threshold cut on the energy proxy. 

%\FloatBarrier

\subsection{The Pierre Auger Observatory and its data sample}

The Pierre Auger Observatory \citep{augernim04,augernim15} (Malarg\"ue, Argentina, 35.2$^\circ$ S, 69.5$^\circ$ W, 1400 m a.s.l.) combines a large surface detector (SD) with an air-fluorescence detector (FD) to observe, in a complementary way, extensive air showers generated by ultrahigh-energy cosmic rays. The SD is composed of an array of 1660 water-Cherenkov stations spread over an area of about 3000 km$^2$. It measures, with a duty cycle of nearly 100\%, the particles (mainly muons, electrons and photons) reaching ground level. The FD, comprising 27 telescopes at four sites, overlooks the surface array and observes, with a duty cycle close to 15\%, the fluorescence light emitted by nitrogen molecules excited by the particles from the air showers.

The data set used in the present work consists of 231 cosmic-ray events with zenith angle $\theta\leq 80^\circ$ and energy $E_{\rm CR}\geq 52$~EeV recorded by the SD from January 1, 2004 up to March 31, 2014 \footnote{The list of events, together with a more detailed description of their selection and reconstruction, can be found in \citep{augerapj2015}.}. These events satisfy a fiducial cut requiring their impact position to be well-contained in the SD array. This ensures an accurate reconstruction of the shower geometry and energy as well as a robust estimation of the exposure. 

The arrival directions of these cosmic rays are determined from the relative arrival times of the shower front in the triggered stations. The angular resolution, defined as the radius around the true cosmic-ray direction that would contain 68\% of the reconstructed shower directions, is better than 0.9$^\circ$ for the energies considered here \citep{augerar}. The energy estimate is obtained from the signals recorded by the SD stations \citep{augerrec1,augerrec2} and is calibrated using `hybrid' events (i.e., detected simultaneously by SD and FD) and using the quasi-calorimetric energy determination obtained with the FD \citep{augerene1,augerene2}. The statistical uncertainty in the energy determination is smaller than 12\% for the energies used here \citep{augerrec2,augerene2}. The systematic uncertainty on the absolute energy scale is 14\%.
 
At energies above 4 EeV the SD trigger is fully efficient, and the determination of the exposure is purely geometrical \citep{augernim10}. For the time period considered and for the applied energy and zenith angle selection, it amounts to 66,452~km$^2$~sr~yr. The adopted zenith-angle selection allows the field of view of the Observatory to extend from $-90^\circ$ to $+45^\circ$ in declination. 

\subsection{Telescope Array and its data sample}

The Telescope Array (TA) is situated in Utah, USA ($39.3^\circ$~N, $112.9^\circ$~W, 1400 m a.s.l.). It consists of 507 plastic scintillator detectors, each of 3~m$^2$ in area,  located on a 1.2~km square grid and covering an area of approximately 700\,km$^2$ (for further details see \cite{AbuZayyad:2012kk}). The atmosphere over the surface array is viewed by 38 fluorescence telescopes arranged in 3 stations. The detector has been fully operational since March 2008.

In this analysis, we use surface-detector data recorded between May 11, 2008 and May 4, 2014.  For the reconstructed events, the energies determined by the SD array were renormalized by 1/1.27 to match the SD energy scale to that of the fluorescence detector, which was determined calorimetrically \cite{taspectrum}.  Of these events, 87 met the following criteria: (1) each event triggered at least four SD counters; (2) the zenith angle of the event arrival direction was less than $55^\circ$; and (3) the reconstructed energy was greater than 57~EeV.  The event-selection criteria described above are the same as in Ref.~\cite{tahotspot} and were optimized to increase the observed number of cosmic rays. The angular resolution of these events is about $1.5^\circ$, while the energy resolution is $\sim 20$\%. The systematic uncertainty on the energy scale is 22\%~\cite{taspectrum}.

To conclude the description of the data, a remark is in order. As already mentioned, the absolute energy scale of UHECRs may contain a systematic error which may be, in principle, different for the two experiments. In accordance with the findings of the TA-Auger Energy Spectrum working group \cite{UHECRMaris}, the UHECR spectra measured by Auger and TA may be made coincident in the region around $10^{19}$~eV (the `ankle' region) by up-shifting the Auger energies (equivalently, down-shifting the TA energies) by $\sim 13$\%. Assuming that the relative shift is energy-independent, and to make the energy scales compatible, we choose to down-shift the TA energies by $13$\% in the likelihood analyses presented here.  Note that assuming lower energies is a conservative choice with respect to the assumed magnetic deflections for the analyses since the deflections grow as the energy decreases. In view of its relatively small magnitude, the overall effect of this shift on the analyses is minor. 

\section{Magnetic deflections of UHECRs}
\label{sec:magdef}

The deflections in the magnetic fields to which, unlike neutrinos, cosmic rays are subject, are crucial in determining the sensitivity of the analyses presented in this paper and any interpretation of the results requires their understanding. The cosmic magnetic fields that deflect the UHECR trajectories are  naturally separated into the Galactic and the extragalactic fields. Of these, the extragalactic fields are the least known. The Faraday rotation measurements of extragalactic sources indicate that the extragalactic fields are smaller than $\sim 10^{-9}$~G \cite{Kronberg:1993vk,Durrer:2013pga,Pshirkov:2015tua}. In this scenario, for a correlation length smaller than $1$~Mpc, the deflections of protons of energy $10^{20}$~eV over a distance of 50~Mpc are smaller than $2^\circ$.

\begin{figure}[t]
\centering
{\includegraphics[width=0.8\linewidth]{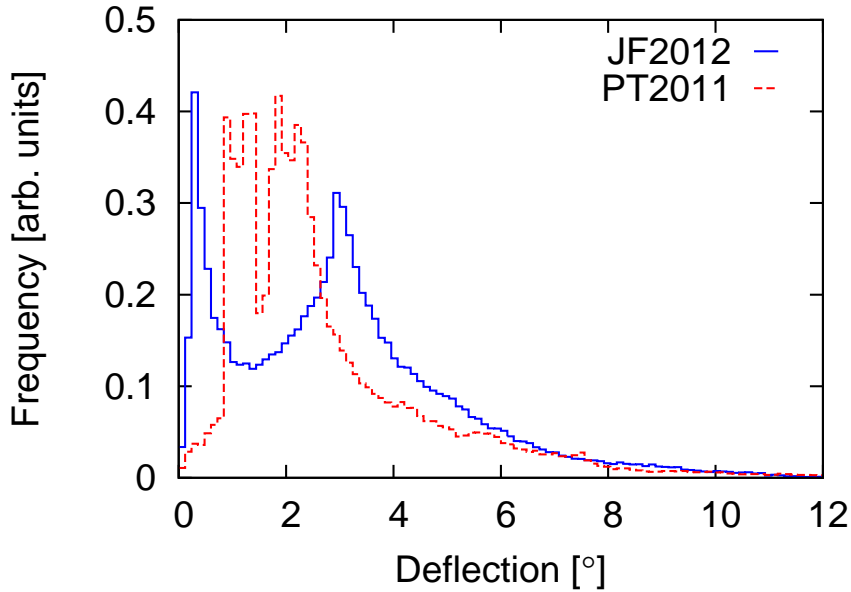}}
\caption{Distribution of UHECR deflections in two Galactic magnetic field models marked PT2011 \cite{Pshirkov:2011um} and JF2012 \cite{Jansson:2012pc} for the regular component. The energies of actual UHECRs are renormalized to show the distributions for $E/Z = 100$~EeV. The double-peak structure is mostly due to the fact that UHECRs from different Galactic hemispheres undergo different deflections.}
\label{fig:deflections}
\end{figure}

The Galactic magnetic field can be further divided into regular and turbulent components. The regular part of the Galactic magnetic field is expected to give a dominant contribution to the UHECR deflections. There are several models of the regular Galactic field in the literature (see, e.g., \cite{Haverkorn:2014jka} for a recent review). While the averages of the magnitudes of the deflections across the sky predicted in these models are similar, individual deflections in a given direction of the sky vary substantially, so that reliable predictions have not yet been achieved. To get an idea of the expected magnitude of the deflections, a set of $10^5$ random cosmic-ray events was simulated following the Auger and TA exposures, weighted in a proportion corresponding to the actual number of events in the data sets of each experiment that we consider in this work. For each of these events we computed the deflection angle according to two recent Galactic magnetic field models \cite{Pshirkov:2011um,
Jansson:2012pc},  assuming the primary CR to be a proton and its energy to be $10^{20}$~eV. The corresponding distributions are shown in Figure~\ref{fig:deflections}. The median deflections from the two models were found to be of the same order, $\sim 3^\circ$ in both cases and having a rather wide distribution. 

Additionally, the cosmic rays are also deflected in the turbulent Galactic magnetic field. There are several estimates of these deflections in the literature~\cite{random-deflections}. In all cases the contribution due to the turbulent fields was found to be sub-dominant as
compared to that due to the regular field. 

Apart from the magnetic field, the magnitude of the deflections also depends on the composition of UHECRs. The current measurements of the UHECR composition at the highest energies are still uncertain \cite{augerxmax,taxmax}, so that the assumed CR deflections are uncertain too.

In summary, it appears unlikely that typical UHECR deflections at $E=10^{20}$~eV are smaller than $2-3^\circ$, but may be significantly larger in the case of a heavier composition. In particular, if the deflections are not too large, they scale approximately as $Z/E$, where~$Z$ is the atomic number. In view of these large uncertainties, when considering likelihood tests we perform the analyses for benchmark deflection values, parameterized by the deflection $D$ of a proton at $E=10^{20}$~eV. For simplicity, we model individual deflections as a random variable with a 2-dimensional Gaussian distribution with the energy-dependent width 
\begin{equation}
\sigma_{\rm MD} (E) = D \times 100\,{\rm EeV}/E. 
\label{eq:sigma_MD}
\end{equation}
We consider test values of $D=3^\circ$, $6^\circ$ and when possible, also of $9^\circ$, which are compatible with the distributions for protons displayed in Figure~\ref{fig:deflections} and may also account for somewhat heavier cosmic rays. 

\section{UHECR correlation analyses with high-energy cascades and high-energy tracks}
\label{sec:UHECR-HESE}

Two different analyses, a cross-correlation test and a stacking likelihood analysis, are performed with the neutrino data sample of 39 high-energy cascade events (described in Section~\ref{sec:HESE_sample}) and the 16 high-energy tracks (7 from the sample described in Section~\ref{sec:HESE_sample} and the 9 described in Section~\ref{sec:HETM_sample}). The total number of UHECRs is $N_{\rm CR}=318$ ($N_{\rm Auger}=231$ and $N_{\rm TA}=87$). For the two analyses, cascades and tracks are considered separately since, due to their different angular resolutions, the angular distance at which a signal (if any) can be observed would be different.  

\subsection{Cross-correlation analysis with high-energy cascades and high-energy tracks}
\label{sec:cross}
If cosmic rays and neutrinos come from the same sources, then the angular separation between their arrival directions would be related to the magnetic deflections suffered by the cosmic rays, convoluted with the resolution in the determination of their respective arrival directions. This angular distance is unknown a priori due to the poor knowledge of the intervening magnetic fields and the uncertain value of the charges of cosmic rays. The cross-correlation method consists of computing the number of UHECR-neutrino pairs as a function of their angular separation $\alpha$, $n_{\rm p}(\alpha)$, and comparing it to the expectation from an isotropic distribution of UHECR arrival directions. The angular scan performed in this case is between $1^\circ$ and $30^\circ$ with a step of $1^\circ$. Due to this scan, the cross-correlation method does not rely on any assumption about the exact value of the strength of the magnetic deflections, unlike the likelihood method described in the following 
subsection (though there is a large trial factor).

To estimate the significance of any excess, the pre-trial $p$-value is first computed by evaluating for each angular distance scanned the fraction of isotropic simulations having more pairs than the data. This is done by keeping the neutrino positions fixed and simulating an isotropic distribution of the arrival directions of UHECRs arriving at the Earth according to the corresponding geometric exposures of the Observatories. Then, we find the angular scale for which this fraction is minimized, the minimum value of this fraction being the pre-trial $p$-value. The post-trial $p$-value is calculated as the fraction of simulations of isotropic arrival directions of cosmic rays that under a similar analysis would give rise to a smaller pre-trial $p$-value than the one observed in the data.

\begin{figure}
  \centering
  \includegraphics[width=0.65\linewidth]{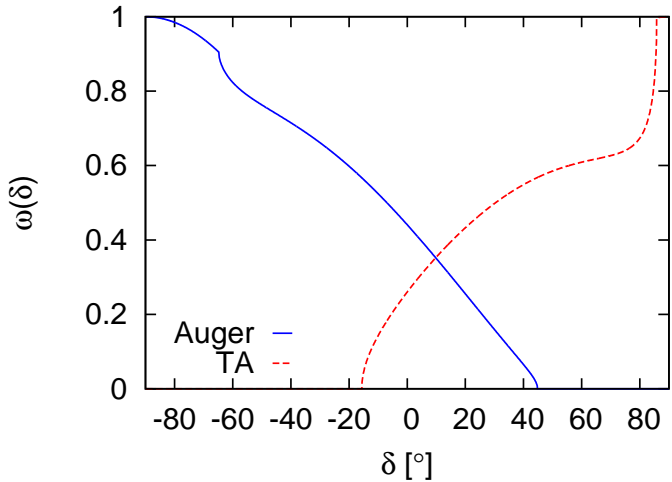}
\caption{The geometrical relative exposures of the Pierre Auger Observatory and Telescope Array as a function of declination. Both distributions have been normalized to unity at the declinations where the exposure is maximum for each experiment. To simulate signal events, the decision whether each event is a cosmic ray measured by Auger or by TA is done with a probability given by the fraction of the number of events measured by each Observatory.}
\label{fig:expo}
\end{figure}

Before applying the cross-correlation method to the data, its sensitivity and discovery potential are computed. The sensitivity is defined as the number of signal events, $n_{{\rm{s}}}$, for which one would obtain a $p$-value of less than 50\%, with respect to the isotropic expectation, in 90\% of the simulations. The discovery potential at, say, 3$\sigma$, is the number of signal events for which one would obtain a $p$-value corresponding to more than $3\sigma$ with 50\% probability in repeated experiments. To obtain these quantities we perform simulations in which a number $n_{{\rm{s}}}$ of cosmic-ray events are assumed to come from sources. They are generated by smearing the positions of the neutrinos with a Gaussian distribution with a standard deviation obtained from the median angular resolution of each neutrino event, while the remaining $N_{{\rm{CR}}}-n_{{\rm{s}}}$ events are assumed to be isotropic. The sources are assumed to have equal apparent luminosity at Earth. To generate the  $n_{{\rm{s}}}$ events, a decision is first made as to whether each event is a cosmic ray measured by Auger or by TA, with a probability given by the fraction of the number of events measured by each Observatory. A source position is then chosen with a probability given by the relative exposure of the corresponding Observatory (shown in Figure~\ref{fig:expo}). 

The arrival direction of the $i^\mathrm{th}$ UHECR event is obtained by smearing the source position with a Gaussian with standard deviation $\sigma(E_i)$ calculated as:

\begin{equation}
\label{eq:sigma}
\sigma(E_i)=\sqrt{\sigma^2_{\rm Auger, TA}+\sigma^2_{\rm MD}(E_i)},
\end{equation}

\noindent where $\sigma_{\rm Auger} = 0.9^\circ$ and $\sigma_{\rm TA} = 1.5^\circ$ are the angular resolutions of Auger and TA, respectively, and $\sigma_{\rm MD}(E_i)$ is the magnetic deflection defined in Eq.~(\ref{eq:sigma_MD}). The UHECR energy $E_{\rm CR}$ is sampled according to the spectrum determined by Auger or TA in the energy range of the UHECRs considered in this analysis, i.e., scaling as $E^{-4.2}$ \cite{Augerindex} and $E^{-4.5}$ \cite{TAindex}, respectively. 

The values obtained for the sensitivity (90\% CL) and for the $3\sigma$ discovery potential are listed in Table~\ref{tab:xcorr} in terms of $n_{{\rm{s}}}$, the total number of cosmic-ray signal events. They imply for instance that for~$D=3^\circ$  a total number of around 15.9 cosmic-ray events should originate on average from the directions of the neutrino tracks (or $93.9$ from the directions of the neutrino cascades) in order to reach a $p$-value corresponding to~$3\sigma$ in 50\% of the cases. Note that since in this case we are considering a total of 318 CR directions, this would correspond to a situation in which $\sim 5$\% ($\sim 30$\%) of the UHECRs are required to have common sources with the neutrino tracks (cascades) in order to reach a $3\sigma$ result if $D=3^\circ$ (somewhat larger fractions are required for the $D=6^\circ$ case). Taking into account that there are 16 tracks and 39 cascades, we see that the average number of required events per source is smaller for the tracks than for the cascades. This is more evident for $D = 3^\circ$ since in this case the signal events happen to lie quite close to the neutrino directions. 

\begin{table}[tbh]
\caption{Sensitivities and post-trial $3\sigma$ discovery potentials DP($3\sigma$) for the cross-correlation method in terms of $n_{\rm s}$, the total number of cosmic-ray signal events. 
}
\label{tab:xcorr}
\begin{center}
\begin{tabular}{c|cc|cc}
\hline
  {}                 & \multicolumn{2}{|c}{High-energy tracks}            &  \multicolumn{2}{|c}{High-energy cascades}\\
  $D$                     & sensitivities &DP($3\sigma$)               & sensitivities  & DP($3\sigma$)\\
  \hline
  $3^\circ$         & 7.0                            & 15.9                                & 43.5                          & 93.9   \\
  $6^\circ$         & 11.1                          & 25.3                                & 50.0                          & 106.1  \\
  \hline
  \end{tabular}
\end{center}  
\end{table}
\FloatBarrier

\subsection{Stacking analysis with high-energy cascades and high-energy tracks}
\label{sec:stack_hese}

Stacking a set of sources is a well-known way of accumulating multiple weaker signals to enhance the discovery potential. Even if the background is also stacked, a better sensitivity can be achieved with this procedure than when looking at the same number of sources separately~\cite{IC40Paper}. Since neutrinos are neither deflected nor absorbed on their way to Earth, here we stack the arrival directions of the neutrinos and search for coincident sources of CRs. 

The unbinned likelihood method is used in this analysis, with the log of the likelihood function defined as:

\begin{equation}
\label{logLLH}
\begin{split}
\log \mathcal L(n_\mathrm{s})=\sum_{i=1}^{N_\mathrm{Auger}}&\log \left ( \frac{n_\mathrm{s}}{N_\mathrm{CR}} \mathcal S^i_\mathrm{Auger} + \frac{N_\mathrm{CR}-n_\mathrm{s}}{N_\mathrm{CR}} \mathcal B^i_\mathrm{Auger} \right)\\
		+ \sum_{i=1}^{N_\mathrm{TA}}&\log \left ( \frac{n_\mathrm{s}}{N_\mathrm{CR}} \mathcal S^i_\mathrm{TA} + \frac{N_\mathrm{CR}-n_\mathrm{s}}{N_\mathrm{CR}} \mathcal B^i_\mathrm{TA} \right),
\end{split}
\end{equation}

\noindent where $n_\mathrm{s}$, the number of signal events, is the only free parameter, $N_\mathrm{CR}=N_\mathrm{Auger}+N_\mathrm{TA}$ is the total number of UHECRs, $S^i_\mathrm{Auger}$ and $S^i_\mathrm{TA}$ are the signal probability density functions (PDFs) for Auger and for TA, respectively, and $B^i_\mathrm{Auger}$ and $B^i_\mathrm{TA}$ are the corresponding background PDFs. The Auger signal PDF has the following form:

\begin{equation}
\label{eq:PAOsigPDF}
\mathcal S^i_\mathrm{Auger}(\vec r_i, E_i) = R_\mathrm{Auger}(\delta_i) \cdot \sum^{N_\mathrm{src}}_{j=1} S_j(\vec r_i,\sigma(E_i)),
\end{equation}

\noindent where $\vec r_i$ is the angular position of the $i^\mathrm{th}$ UHECR event, $R_\mathrm{Auger}(\delta_i)$ takes into account the Auger detector response, e.g., the relative exposure for given event declination $\delta_i$ (see Figure~\ref{fig:expo}) and $N_\mathrm{src}$ is the number of stacked sources, 39 for the cascades and 16 for the tracks. The last term, $S_j(\vec r_i,\sigma(E_i))$ is the value of the normalized directional likelihood map for the $j^\mathrm{th}$ source (i.e., the $j^\mathrm{th}$ neutrino) taken at $\vec r_i$ and smeared with a Gaussian with standard deviation $\sigma(E_i)$, defined in Eq.~\ref{eq:sigma}, with $\sigma_\mathrm{Auger}=0.9^{\circ}$.

The signal PDF for Telescope Array $S^i_\mathrm{TA}$ has the same form as Eq.~\ref{eq:PAOsigPDF}, but the relevant parts are replaced with the Telescope Array equivalents, namely $R_\mathrm{Auger}(\delta_i)$ is replaced by the Telescope Array relative exposure $R_\mathrm{TA}(\delta_i)$ (see Figure~\ref{fig:expo}) and the angular resolution is $\sigma_\mathrm{TA}=1.5^{\circ}$.

The background PDFs, $B^i_\mathrm{Auger}$ and $B^i_\mathrm{TA}$, represent the probabilities of observing a cosmic ray from a given direction assuming an isotropic flux. Therefore they are taken to be the Auger and TA normalized exposures (see Figure~\ref{fig:expo}).

\begin{figure}[!h]
\begin{center}
  \centering
  \subfigure[]{\includegraphics[width=0.49\linewidth]{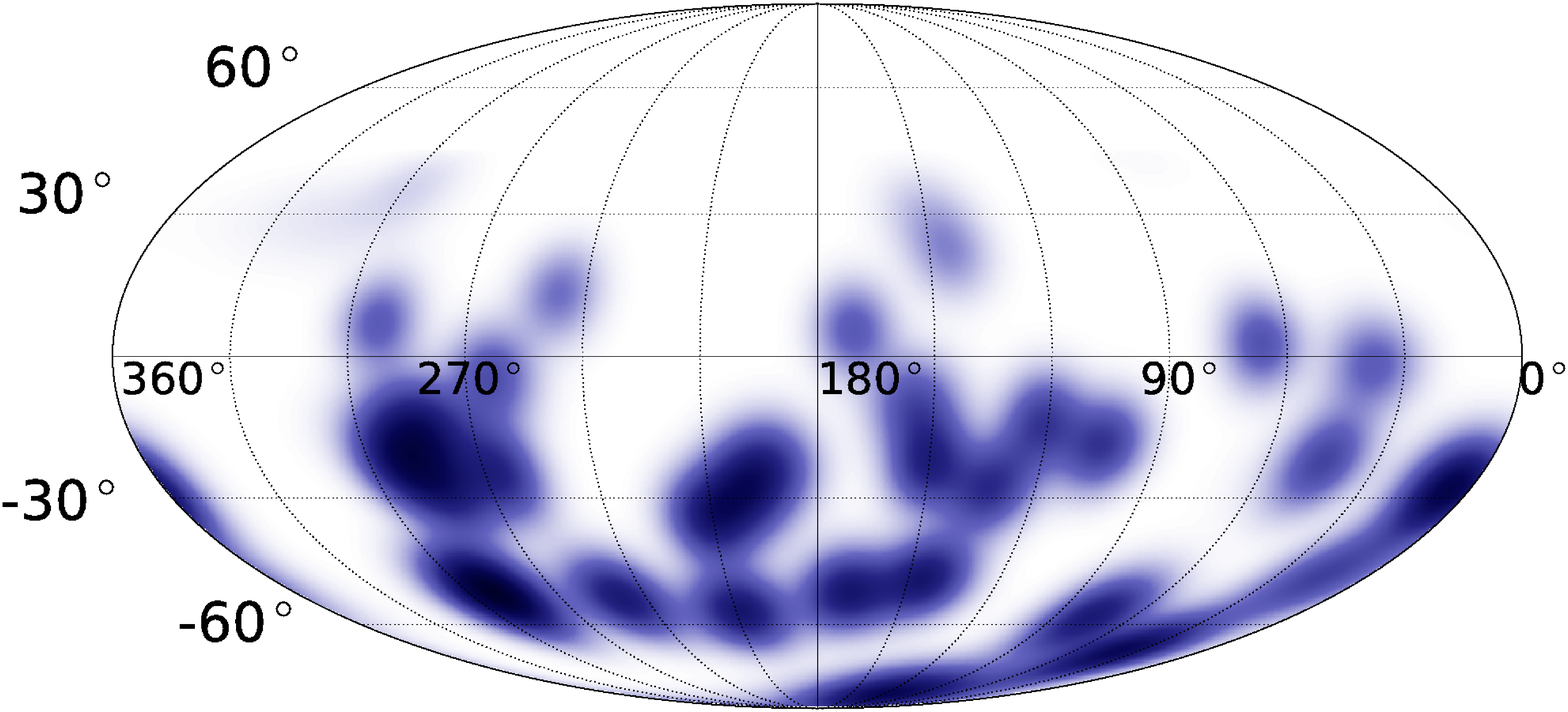}}
  \subfigure[]{\includegraphics[width=0.49\linewidth]{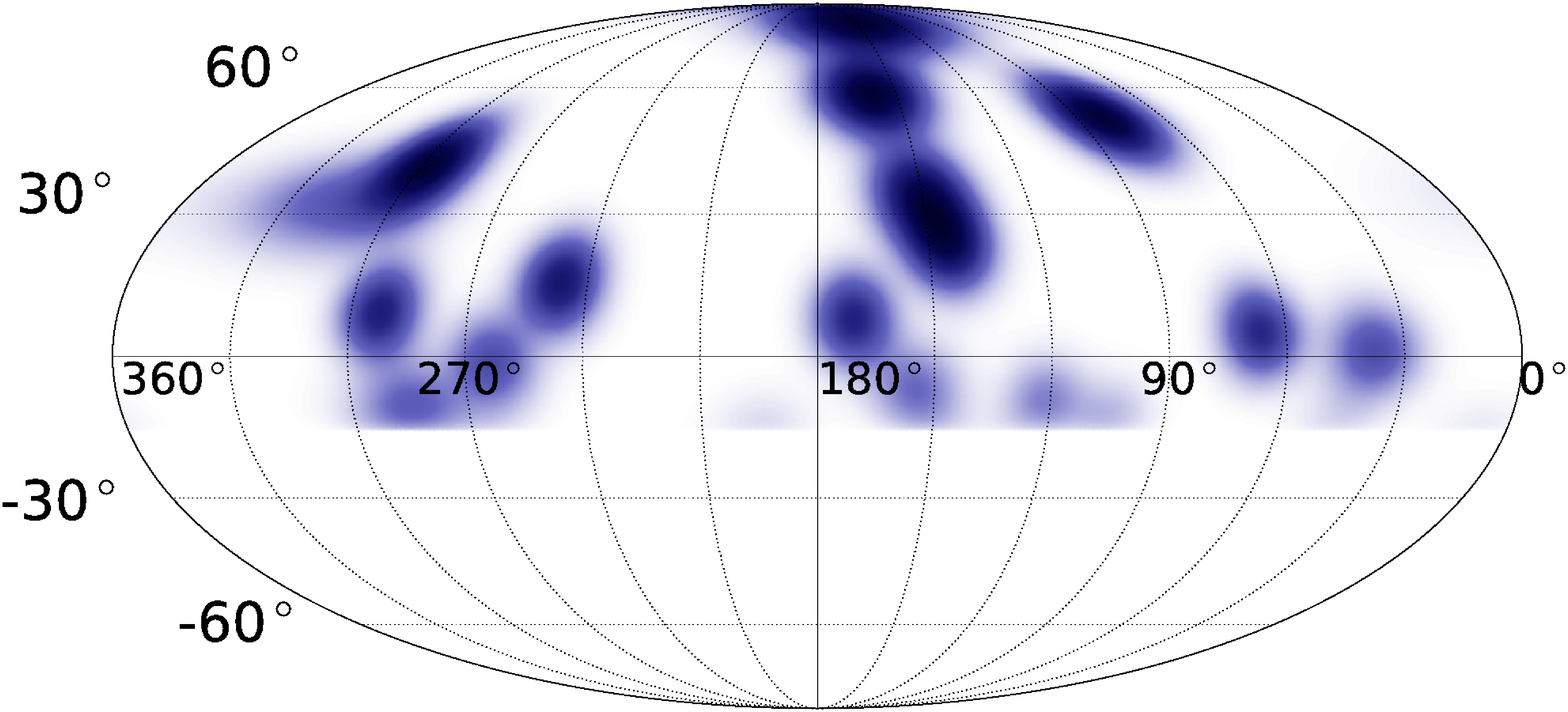}}
  \subfigure[]{\includegraphics[width=0.49\linewidth]{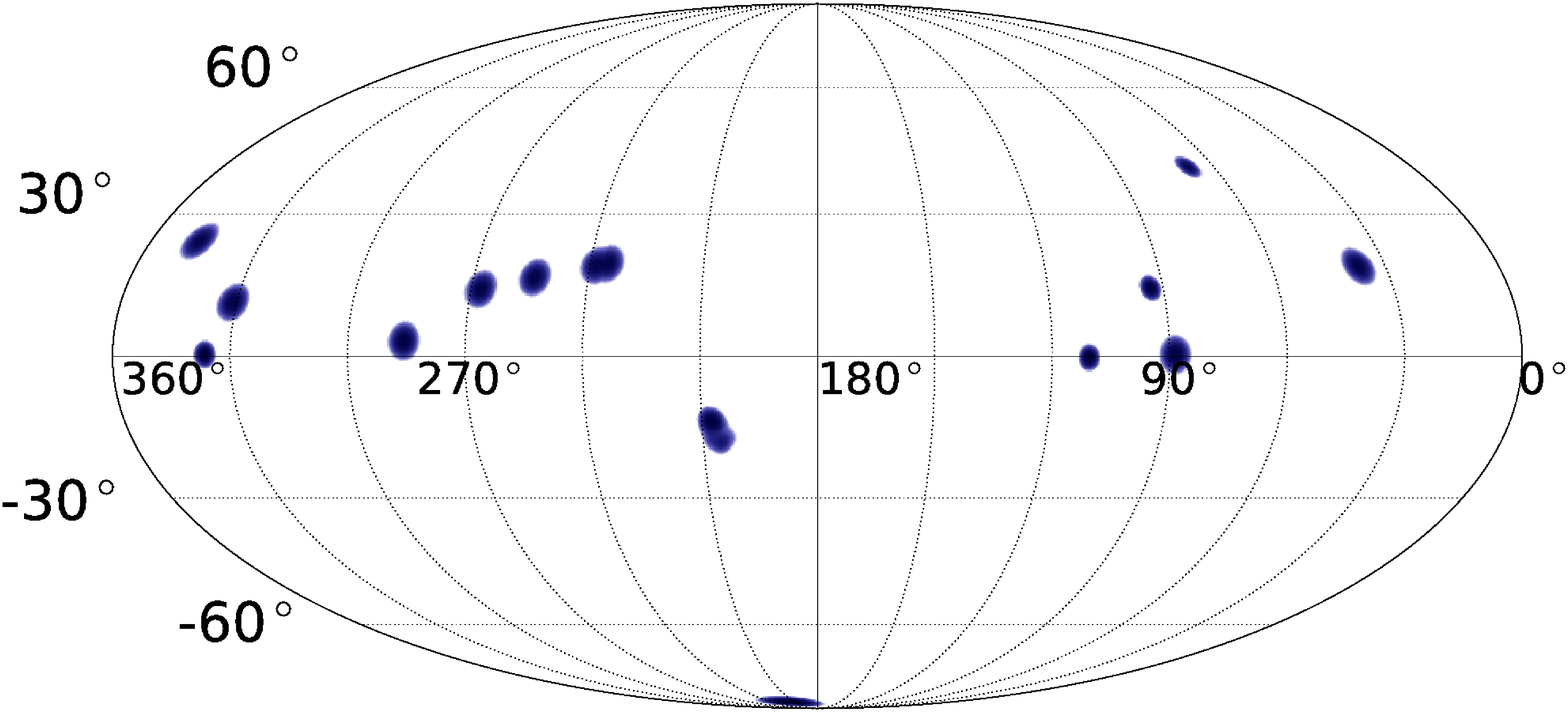}}
  \subfigure[]{\includegraphics[width=0.49\linewidth]{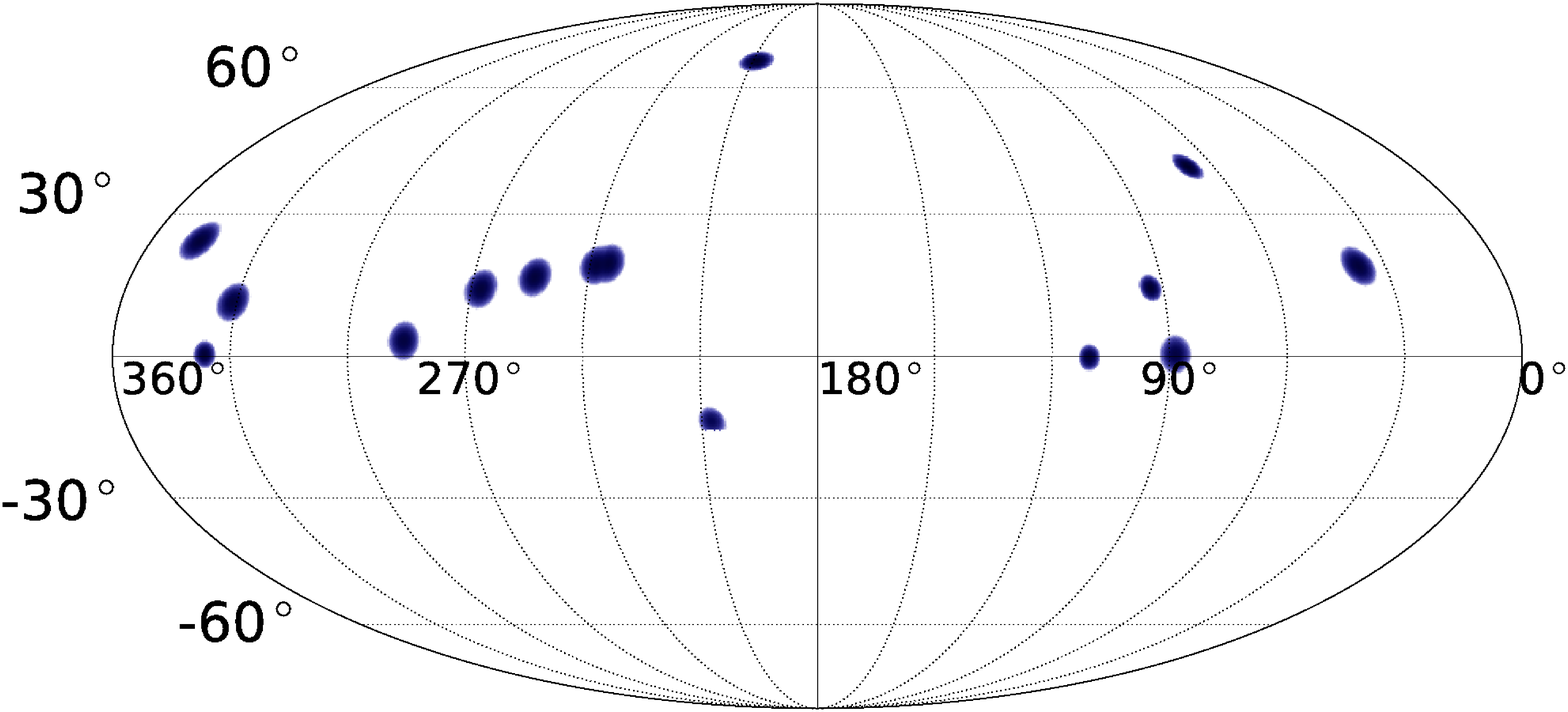}}
  \caption{Neutrino likelihood maps convoluted with the exposures of the CR Observatories in Equatorial coordinates.
The two upper plots are for the high-energy cascades, while the two lower ones are for the high-energy tracks. The declination-dependent exposure of Auger is applied in (a) and (c) and that of TA in plots (b) and (d). }
  \label{fig:signal_pdf}
\end{center}
\end{figure}

In Figure~\ref{fig:signal_pdf} we show the normalized neutrino likelihood maps convoluted with the exposures of the two CR Observatories, to demonstrate the spread of the different observed neutrino arrival directions visible from each site.

%\pagestyle{empty}
%\newgeometry{top=1.3cm}
%\restoregeometry
%\clearpage
%\pagestyle{plain}

The test statistic $TS$ is defined as:
\begin{equation}
TS=2 \log \frac{\mathcal L(n_\mathrm{s})}{\mathcal L(n_\mathrm{s}=0)}.
\end{equation}

In accordance with Wilk's theorem \cite{Wilkstheorem}, the test statistic is expected to follow a distribution close to $\chi^2$ of one degree of freedom, as $n_{\mathrm{s}}$ is the only free parameter. This assumption has been explicitly verified and is used to calculate the pre-trial $p$-values.

To evaluate the sensitivity and discovery potential of this method we perform the same simulation of cosmic-ray signal events as in Section 4.1. The only difference is  that in this case the sources are sampled from each neutrino likelihood map.
In Table~\ref{tab:disco_stack} the discovery potential and sensitivity calculations are shown in terms of the total number of cosmic-ray signal events.
We note that in these simulations we adopted the same value of deflection parameter $D$ in the ${\log \mathcal L}$ as the one used to simulate the CR deflections. If the actual deflections were different than the values of $D$ considered in ${\log \mathcal L}$ the discovery potential would become worse.
The comparison of the discovery potentials obtained for this method with those obtained for the cross-correlation in the previous subsection is non-trivial, because in the cross-correlation case they are already penalized for the trial factor of the scan in the angular scale. On the other hand, for the likelihood method a fixed deflection parameter $D$ is adopted, and hence smaller required values of $n_{\rm s}$ may result with the likelihood method.

\begin{table}[tbh]
\caption{Sensitivities and pre-trial $3\sigma$ and $5\sigma$ discovery potentials (DP($3\sigma$) and DP($5\sigma$)) for the likelihood stacking method in terms of the total number of cosmic-ray signal events.}
\label{tab:disco_stack}
\begin{center}
\begin{tabular}{c|ccc|ccc}
\hline
                  & \multicolumn{3}{|c}{High-energy tracks}                        &  \multicolumn{3}{|c}{High-energy cascades}\\
  {$D$ }         & sensitivities   &DP($3\sigma$)    &DP($5\sigma$)          & sensitivities & DP($3\sigma$)&DP($5\sigma$) \\
  \hline
  $3^\circ$  & 7.4           & 15.7        &  29.1       & 22.8          & 55.4   & 89.9  \\
  $6^\circ$  & 12.9          & 27.1        &  47.8       & 32.0          & 72.6   & 121.2 \\
  $9^\circ$  & 18.0          & 38.5        &  64.5       & 53.1          & 98.6  & 163.7 \\
  \hline
  \end{tabular}
  \end{center}
\end{table}

\FloatBarrier

\subsection{Results}

\begin{figure}[!h]
   \centering
%  \subfigure[]
 {\includegraphics[width=0.49\linewidth]{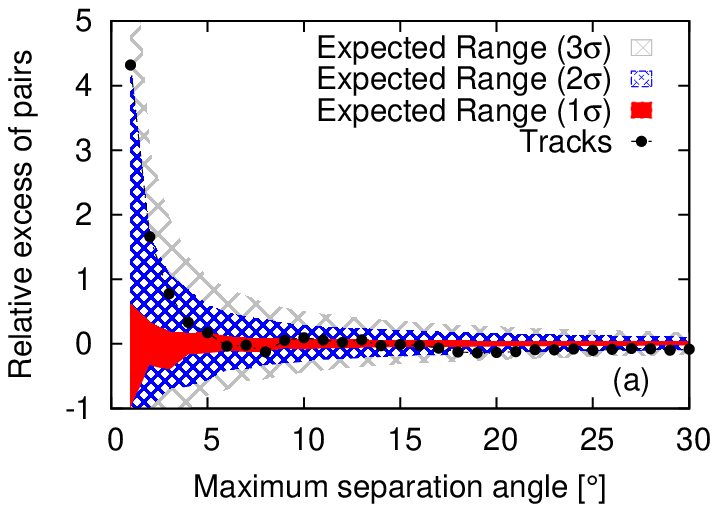} \label{cctracks}}
%  \subfigure[]
 {\includegraphics[width=0.49\linewidth]{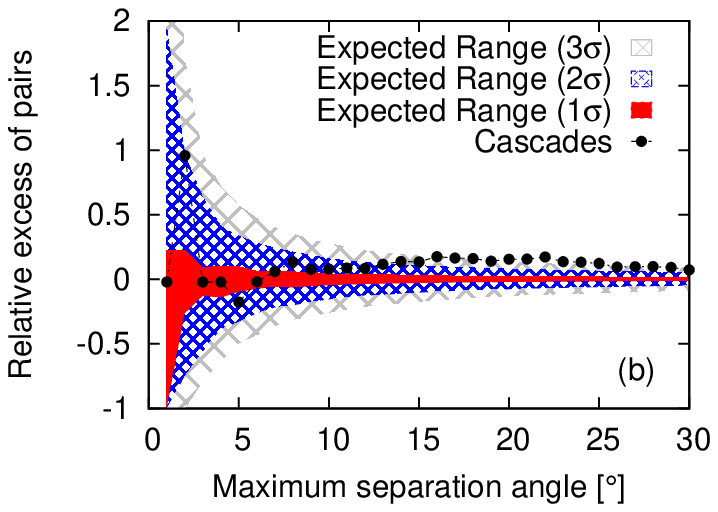} \label{cccascades}}
   \caption{Relative excess of pairs, $[n_{\rm p}(\alpha)/ \langle n_{\rm p}^{\rm{iso}}(\alpha) \rangle ]-1$, as a function of the maximum angular separation between the neutrino and UHECR pairs, for the analysis done with the track-like events (a) and with the cascade events (b). The $1\sigma$, $2\sigma$ and $3\sigma$ fluctuations expected from an isotropic distribution of arrival directions of CRs are shown in red, blue and grey, respectively.} \label{cc}
 \end{figure}

The results of the cross-correlation method applied to the data are shown in Fi\-gu\-res~\ref{cc}a and~\ref{cc}b for the high-energy tracks and high-energy cascades, respectively. Here the fractional excess relative to the expectation for an isotropic CR distribution, $[n_{\rm p}(\alpha)/ \langle n_{\rm p}^{\rm{iso}}(\alpha) \rangle ]-1$, is plotted as a function of the angular separation between the neutrino and cosmic-ray pairs. The dots represent the data, and the ranges corresponding to the $1\sigma$, $2\sigma$ and $3\sigma$ confidence bands obtained in isotropic realizations of CR arrival directions are also shown.

For the cross-correlation with the sample of neutrino tracks, the maximum departure from the expectation of an isotropic CR flux occurs at an angular distance of $1^\circ$ (Figure~\ref{cc}a), where $0.38$ pairs were expected on average and 2 pairs are detected. The post-trial $p$-value is $28\%$. For the analysis done using the high-energy cascades (Figure~\ref{cc}b), the smallest pre-trial $p$-value is obtained at an angular distance of $22^\circ$, for which 575 pairs are observed while 490.3 were expected on average. The post-trial $p$-value is $5.0 \times 10^{-4}$ assuming an isotropic flux of CRs arriving at the Earth.

\begin{figure}[!h]
  \centering
%  \subfigure[]
{\includegraphics[width=\linewidth]{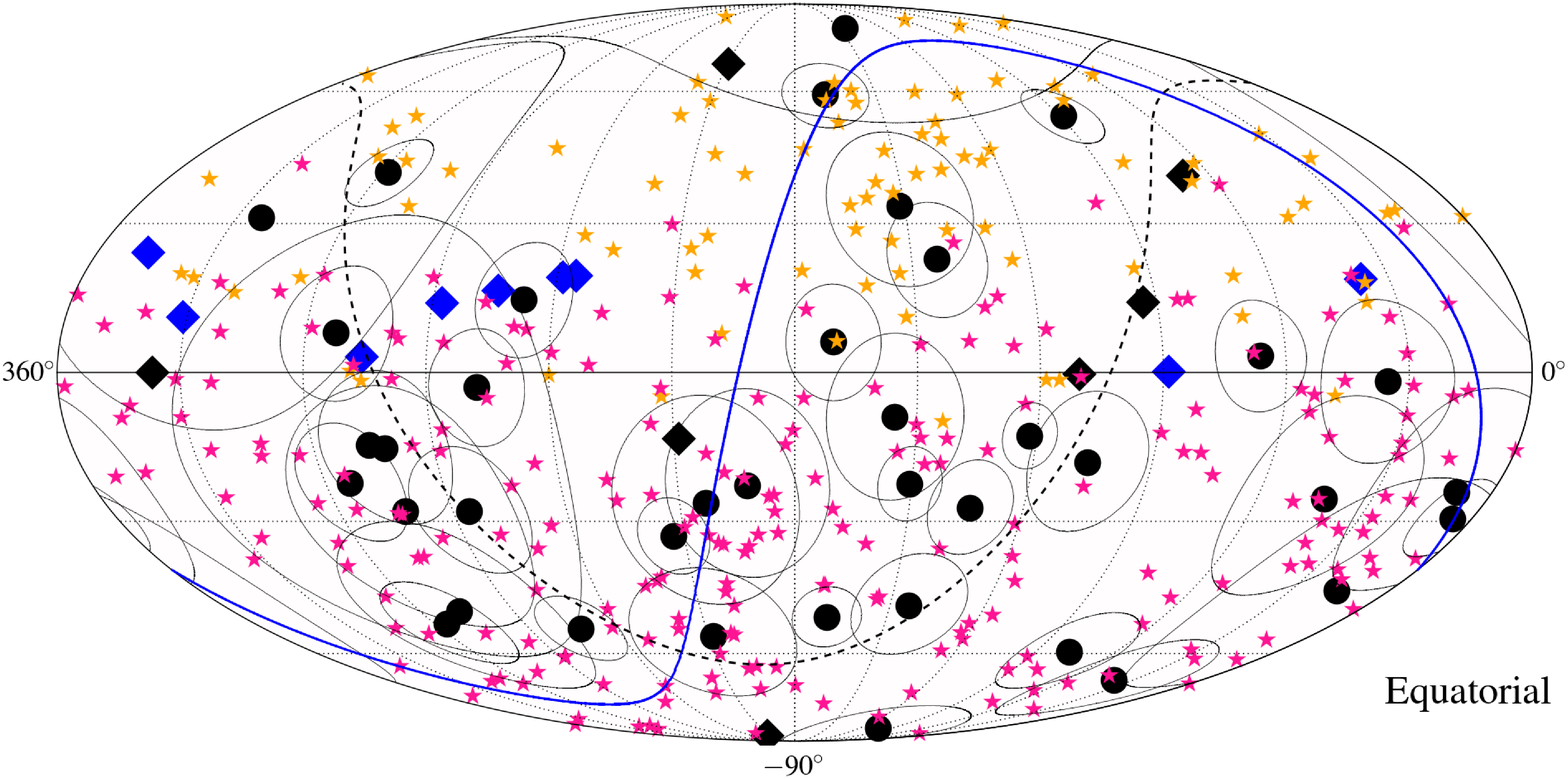}}
{\includegraphics[width=\linewidth]{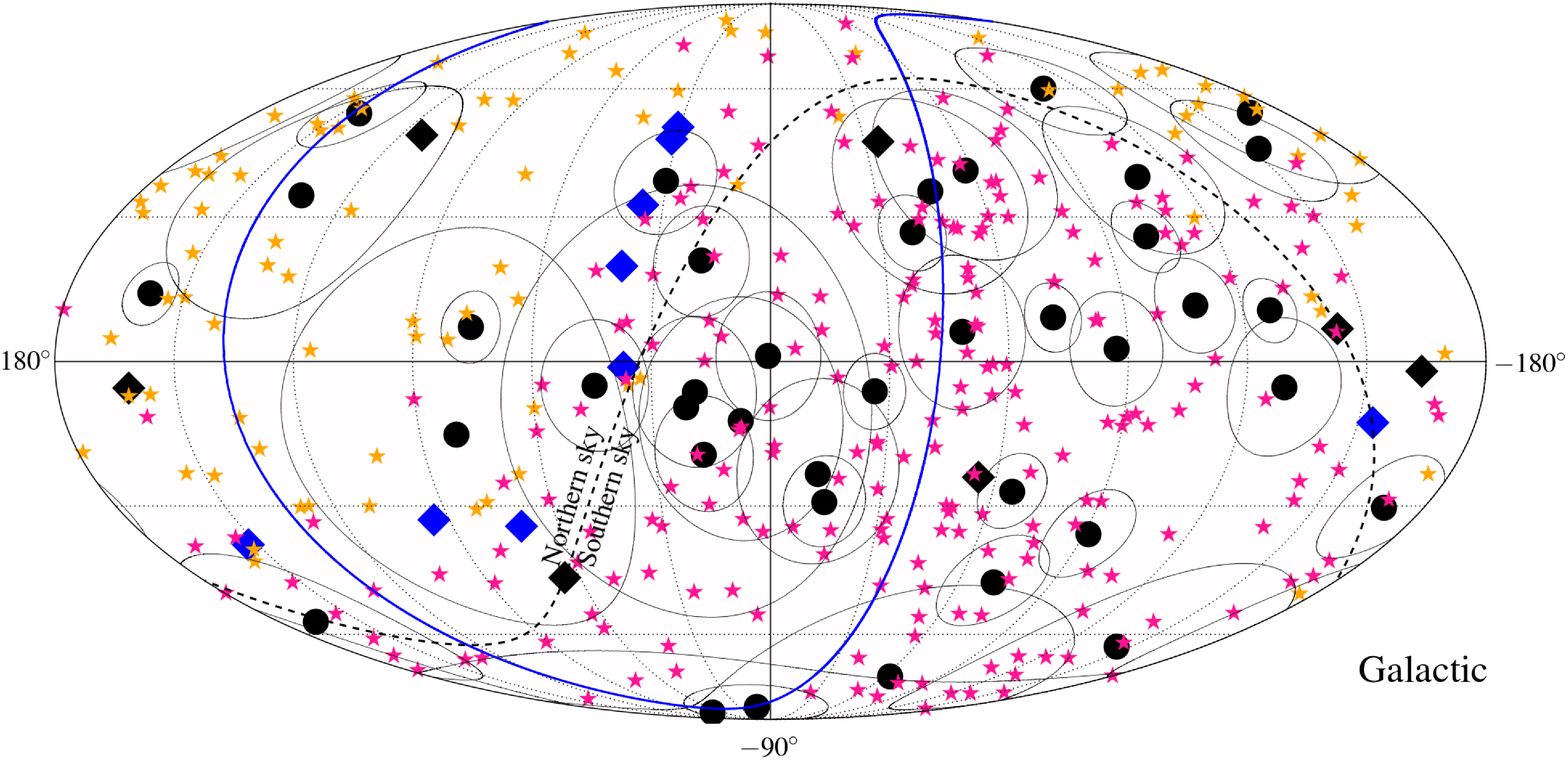}}
%  \subfigure[]
  \caption{Maps in Equatorial and Galactic coordinates showing the arrival directions of the IceCube cascades (black dots) and tracks (diamonds), as well as those of the UHECRs detected by the Pierre Auger Observatory (magenta stars) and Telescope Array (orange stars). The circles around the showers indicate angular errors. The black diamonds are the HESE tracks while the blue diamonds stand for the tracks from the through-going muon sample. The blue curve indicates the Super-Galactic plane.} \label{map_all}
\end{figure}

By looking at Figure~\ref{map_all} one can infer that most of the excess of pairs in this case is due to the fact that there are several high-energy cascades in regions with large densities of UHECRs, i.e., near the Super-Galactic plane and at the TA  `hot spot'~\cite{tahotspot}, which is a~$20^\circ$ radius region centered at the Galactic coordinates $(\ell,b)\simeq(177^\circ,50^\circ$)\footnote{The presence of two cascade neutrino events near this hot spot was already pointed out in Ref.~\cite{olinto}.}. 

We also perform some a posteriori tests of the cross-correlation results, considering separately the data of Auger and TA. It was observed that both samples lead to a minimum at $22^\circ$, with the TA post-trial $p$-value being $9.3 \times 10^{-4}$ and that of Auger being $4.1 \times 10^{-2}$. Thus, when considering the entire UHECR data set, this minimum gets reinforced. 

The results of the likelihood stacking method are summarized in Table~\ref{tab:HESEstackingRezTr}. The most significant deviation from the isotropic flux is found for the magnetic deflection parameter $D=6^{\circ}$ for the cascade sample. The observed pre-trial $p$-value is $2.7\times 10^{-4}$. Due to this rather small value the post-trial $p$-value calculation based on generating background-only samples and counting the fraction of those more significant than the result is not feasible. We then conservatively apply a trial factor of 3 to account for the 3 values of the magnetic deflection parameter $D$ used in the analysis\footnote{This approach is conservative since when using generated background-only samples it was observed that the significances obtained for $D=3^{\circ}$, $6^{\circ}$, and $9^{\circ}$ are strongly correlated. When these simulations were used to obtain trial factors for less significant pre-trial $p$-values we obtained trial factor values smaller than 2.}. The obtained post-trial $p$-value is $8.0\times 10^{-4}$. 

It is important to stress that all the $p$-values quoted for both analyses above are with respect to the null hypothesis of an isotropic UHECR flux, as analyses of the distributions of their arrival directions yielded no evidence of anisotropy at discovery level. However, directions with higher densities of UHECRs, such as the TA `hot spot'~\cite{tahotspot} and the direction of Cen A~\cite{augerapj2015}, have been reported. Hence, as an additional a posteriori study for both analyses, we have also evaluated the significance under the hypothesis of an isotropic distribution of neutrinos. In this case, the UHECR positions have been kept fixed, thus preserving the degree of anisotropy in the arrival directions of CRs. The arrival directions of neutrinos have been simulated producing random right ascensions, while keeping their declination fixed and thus accounting for the declination dependence in the IceCube acceptance. This random-neutrino test assesses the extent to which the $p$-values previously obtained for the clustering of cosmic rays around the neutrino directions could be affected by chance alignments of the neutrinos with known clustering in the cosmic-ray sky.

For the cross-correlation analysis the post-trial $p$-value obtained under the hypothesis of an isotropic distribution of neutrinos is $8.5 \times 10^{-3}$. A similar post-trial $p$-value can be calculated for the likelihood stacking analysis by applying the analysis for all three angular deflection hypothesis $D=3^{\circ}$, $6^{\circ}$, and $9^{\circ}$ and selecting the most significant result. This can then be compared to the significance obtained for real data and it was found that in four cases out of 3000 the significance of the generated samples was higher. Thus, for the likelihood stacking analysis the post-trial $p$-value with respect to the isotropic neutrino flux hypothesis is $1.3\times10^{-3}$ (i.e., of about 3$\sigma$). We see that for both the cross-correlation and the likelihood stacking analyses, the $p$-values obtained under the null hypothesis of isotropic neutrinos turn out to be larger than the ones obtained under the null hypothesis of isotropic CRs, the differences reflecting the extent to which the original $p$-values, from the isotropic cosmic-ray hypothesis, are due to an accidental alignment of the neutrinos with the known clustering of the cosmic rays.

\begin{table}
\begin{center}
\begin{tabular}{l|lll|lll}
\hline
  & \multicolumn{3}{|c}{High-energy tracks}                        &  \multicolumn{3}{|c}{High-energy cascades}\\
 $D$ & $n_\mathrm{s}$ & $TS$ & pre-trial $p$-value& $n_\mathrm{s}$ & $TS$ & pre-trial $p$-value\\
 \hline
 $3^{\circ}$ &  4.2   & 0.6                  &  0.22              & 53.7  & 8.21  & $2.1\times 10^{-3}$\\
 $6^{\circ}$ &  0.5   &  $2.7\times 10^{-3}$ &  0.48              & 85.7  & 11.99 & $2.7\times 10^{-4}$\\
 $9^{\circ}$ &  0     &  0                   &  under-fluctuation & 106.1 & 11.32 & $3.8\times 10^{-4}$\\
 \hline
 \end{tabular}
\end{center}
\caption{Results for the likelihood stacking analyses with the high-energy tracks and high-energy cascades.}
\label{tab:HESEstackingRezTr}
\end{table}

\begin{figure}[!h]
  \centering
  \includegraphics[width=3.5in]{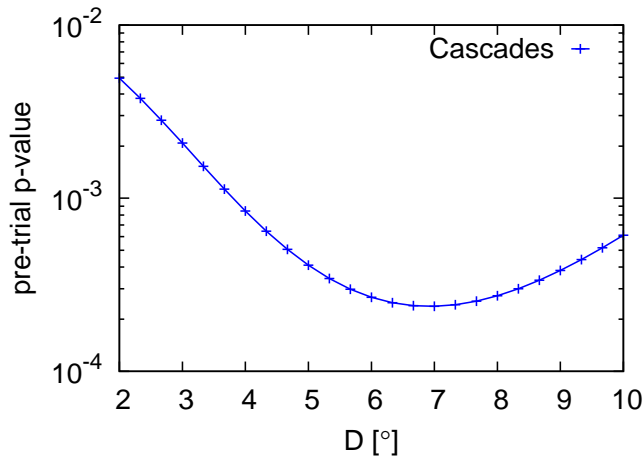}
  \caption{A posteriori angular scan for the stacking with the high-energy cascades. The analysis was done only for the three chosen values of $D$: $3^{\circ}$, $6^{\circ}$ and $9^{\circ}$.}
  \label{fig:apAngScan}
\end{figure}

An a posteriori scan over the values of the assumed deflection $D$ for the stacking analysis with cascades, shown in Figure~\ref{fig:apAngScan}, reveals that the minimum $p$-value happens very close to one of the values adopted for the analysis, i.e., $D=6^{\circ}$.

The angular distance at which an excess would occur in the case of the cross-correlation includes not only the magnetic deflections at the corresponding CR energies but also the experimental angular uncertainties. In the case of cascades, the angular uncertainty is $\sim15^\circ$ and it is $\sim1^\circ$ for the CRs. Since most CRs in the data set have $E_{\rm CR} \sim60$\,EeV, the assumed magnetic deflection, where the smallest $p$-value is found in the case of the likelihood analysis with the cascades ($\sigma_{\rm MD} (E_{\rm CR}) = 6^\circ \times 100\,{\rm EeV}/E_{\rm CR}$), is $\sim10^\circ$ in most cases. To translate this into an angular scale where one would find an excess in the cross-correlation analysis (if there were a signal), we add the different angular scales in quadrature, obtaining $\sqrt{(15^\circ)^2+(1^\circ)^2+(10^\circ)^2} \approx 18^\circ$.  This scale is comparable to the $22^\circ$ where the smallest $p$-value is found for the cross-correlation performed with the cascades. Hence, the 
magnetic deflection of the CRs one would infer from the cross-correlation analysis with the cascades is comparable to the value leading to the smallest $p$-value in the likelihood analysis, even if none of the results are at a level where any strong claims can be made.

The consistent and significant correlation of UHECRs with the high-energy cascades observed in both the cross-correlation and stacking analyses forms a potentially interesting result, which we will continue to monitor in the future.

\section{Stacking search with the 4 year point-source sample}
\label{sec:pssample}
\subsection{Method and discovery potential}

This analysis looks for excesses in the neutrino point-source data set, consisting of through-going tracks and described in Section~\ref{4pssamle}, around the directions of the highest-energy UHECRs. We use a likelihood method which is a generalization of the one described in Section~\ref{sec:stack_hese} \cite{IC86PSPaper,method}, but now the measured positions of the UHECRs are stacked as sources.

The log of the likelihood function is defined as:

\begin{equation}
\label{eq:likelihood}
\log \mathcal{L}(n_{\rm s}, \gamma) = \sum\limits_{i=1}^{N_\nu} \log \left(\frac{n_{\rm s}}{N_\nu}\mathcal{S}_i^{\rm tot} + \left (1-\frac{n_{\rm s}}{N_\nu} \right)\mathcal{B}_i\right),
\end{equation}
 where $n_{\rm s}$ is the number of signal events in the sample and $\gamma$ is the spectral index of the neutrino source candidates, assumed to collectively follow an unbroken power-law spectrum $\propto E^{-\gamma}$. $N_\nu$ is the total number of astrophysical neutrino candidate events in the sample. $\mathcal{S}_i^{\rm tot}$ is the signal PDF for the stacked sources and $\mathcal{B}_i$ is the background PDF.

Signal events from an astrophysical source $j$ are expected to be clustered around the direction of the source $\vec{r}_{j}$ according to the Gaussian distribution:

\begin{equation}
\label{eq:gaussianspatial}
G_i^j = \frac{1}{2\pi\sigma^2_{ij}}
\exp\left(-\frac{|\vec{r}_i-\vec{r}_{j}|^2}{2\sigma^2_{ij}}\right), 
\end{equation}
 
\noindent where $\sigma_{ij}=\sqrt{\sigma_i^2+\sigma_j^2+\sigma^2_{\rm MD}(E_j)}$ accounts for the spread due to the angular resolution of the IceCube event $i$, that of the CR event $j$, as well as for the spread due to  the assumed magnetic deflection of the CR with energy $E_j$.

The signal PDF  $S_i^j$ also accounts for the energy- (and declination-) dependent  response of the IceCube detector, $P(\delta_i,E_i|\gamma)$, which can be obtained from Monte Carlo simulations of events produced by a source with a power-law spectrum of index $\gamma$. The single-source signal PDF is then
\begin{equation}
\label{eq:sigpdf}
\mathcal{S}(\vec{r}_i,\vec{r_j},E_i,\gamma) = G_i^j \times P(\delta_i,E_i|\gamma).
\end{equation}
The total PDF obtained after stacking the $N_{\rm CR}$ UHECR directions is given by
\begin{equation}
\label{eq:stackingspdf}
\mathcal{S}_i^{\rm tot} = \frac{\sum\limits_{j=1}^{N_{\rm CR}} R_\mathrm{IC}(\delta_j,\gamma)\mathcal{S}(\vec{r}_i,\vec{r}_j,E_i,\gamma)}{\sum\limits_{j=1}^{N_{\rm CR}} R_\mathrm{IC}(\delta_j,\gamma)},
\end{equation}

\noindent where $R_\mathrm{IC}(\delta_j,\gamma)$ is the detector acceptance of IceCube at declination $\delta_j$ for a source of spectral index $\gamma$.

Events which are atmospheric muons and/or neutrinos are expected to be distributed uniformly within each declination band. The distribution of the values of the energy estimator $E$ of these events can be indicated by $P(\delta_i,E|\phi_{\rm atm})$. The background probability distribution can then be expressed as
\begin{equation}
\label{eq:bkgpdf}
\mathcal{B}(\vec{r}_i,E_i) = B(\delta_i)\times P(\delta_i,E_i|\phi_{\rm atm}),
\end{equation}
 
\noindent where $B(\delta_i)$ is the declination dependence of the sample.

The number of signal events $n_{\rm s}$ and the spectral index  of the sources $\gamma$ are unknown, but the best estimates of these two parameters,  $\hat{n}_s$ and  $\hat{\gamma}$, can be obtained as those maximizing the likelihood. The test statistic is then obtained as
\begin{equation}
\label{eq:teststat2}
TS = 2 \log\frac{\mathcal{L}(\hat{n}_{\rm s}, \hat{\gamma})}{\mathcal{L}(n_{\rm s}=0)}.
\end{equation}
The significance of an observation can then be estimated by repeating the process on datasets randomized in right ascension  and computing the fraction of randomized samples that produce a value of the $TS$ bigger than that observed in the data. 

The performance of this kind of search can be quantified by the median source flux required for 5$\sigma$ discovery, defined similarly to that in Section \ref{sec:stack_hese} but in terms of the astrophysical neutrino flux. This is calculated by repeating the test on datasets with $n_{\rm s}$ additional simulated signal events injected in the direction of the sources.

Since using all of the 318 events described in Section~\ref{sec:datasets} as sources with their associated angular extensions would essentially cover the whole sky,  the possible gain associated to the stacking of more sources is reduced. Restricting the UHECR sample to those with energies above a given threshold $E_{\rm th}$ could give the advantage of having smaller associated magnetic spreads, potentially enhancing the discovery potential. We look for an optimum value of $E_{\rm th}$ for which the per-source flux required for discovery can be minimized as a consequence of these two competing effects.

\begin{figure}[ht]\centering
\includegraphics[width=0.8\linewidth]{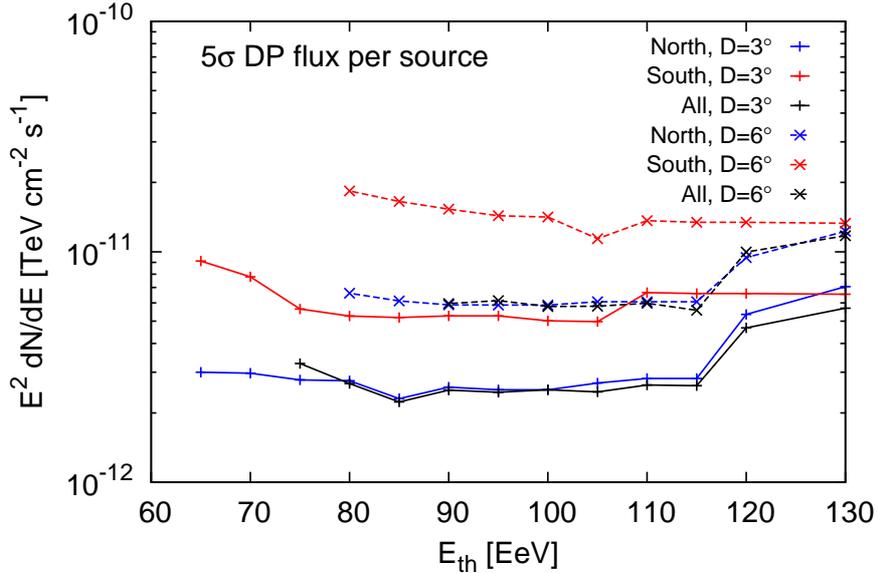}
\caption{ Pre-trial discovery potential  at the $5\sigma$ level. The  normalization of the flux per source required is shown as a function of the UHECR threshold energy $E_{\rm th}$. The results from Northern and Southern skies, as well as the whole sky, are shown for $D=3^\circ$ and $6^\circ$. }
\label{fig:psstackingdisco}
\end{figure}

A simulation was then performed to determine the discovery potential of the UHECR stacking analysis obtained for different values of $E_{\rm th}$ (see Figure~\ref{fig:psstackingdisco}) for both the Southern and Northern hemispheres separately as well as for the whole sky sample. To calculate the discovery potentials, events were injected from point sources in the direction of the UHECRs,  adopting source spectra with $\gamma=2$ and considering two deflection hypotheses, $D = 3^{\circ}$ and $6^{\circ}$. Each point source was positioned with a deflection $\Delta_j$ from the UHECR, where $\Delta_j$ is sampled randomly from a Gaussian distribution of width $\sqrt{\sigma_j^2+\sigma^2_{\rm MD}(E_j)}$. As seen in Figure~\ref{fig:psstackingdisco}, the discovery potentials are characterized by a broad flat region, based on which a value of $E_{\rm th} =  85$\,EeV was chosen. This keeps the 12 highest-energy UHECRs in the Southern sky and 15 in the Northern sky.

Due to the different energy ranges between the neutrino candidate events in the Southern hemisphere ($\sim$100 TeV - 100 PeV) and in the
Northern hemisphere ($\sim$1 TeV - 1 PeV), the flux normalization required for Northern sources turns out to be smaller than for Southern ones. 
However, thanks to the gain provided by having more stacked sources, the whole sky sensitivity turns out to be similar to the one in the Northern sky alone (Figure \ref{fig:psstackingdisco}). We have thus considered just the whole sky sample with  $E_{\rm th} =  85$\,EeV in this analysis.

\subsection{Results}

Applying this analysis to the data, all observations were found to be compatible with the background hypothesis. For the deflection hypothesis of $D = 3^{\circ}$, the $p$-value was found to be 17.3\% with a fitted excess of $n_{\rm s} \sim 123$ events and a fitted source spectrum of $E^{-3.24}$, while the $p$-value was found to be $>50\%$ for the $D = 6^{\circ}$ hypothesis. Accounting for the trial factor due to having tested both the $D = 3^{\circ}$ and the $D = 6^{\circ}$ hypotheses, the post-trial $p$-value was found to be 25.4\%.

\section{Conclusions}
\label{sec:concl}

Three analyses have been performed to investigate correlations between the 318 UHECRs measured by the Auger Observatory and the Telescope Array with various samples of IceCube neutrino events, and the results obtained are all below $3.3\sigma$.

The smallest post-trial $p$-values were obtained when considering IceCube high-energy cascades. In the case of the cross-correlation analysis, the minimum was found for an angular scale of $22^\circ$, with a post-trial $p$-value of $5.0 \times 10^{-4}$ under the assumption of an isotropic flux of UHECRs and $8.5 \times 10^{-3}$ in the a posteriori study under the assumption of an isotropic flux of neutrinos. For the stacking done with the high-energy cascades and a deflection hypothesis of $D = 6^{\circ}$, a  post-trial $p$-value of $8.0\times 10^{-4}$ was obtained under the assumption of an isotropic flux of UHECRs and $1.3\times10^{-3}$ under the a posteriori assumption of an isotropic flux of neutrinos. The two results correspond to comparable magnetic deflections of CRs. These correlation excesses arise mostly from pairs of events in the region of the TA hot spot and also in regions close to the Super-Galactic plane where some excess of events is seen in the Auger sample, but 
at present the results are not significant enough to draw any strong conclusions and are also compatible with being fluctuations of an isotropic distribution. It will be interesting to further study these findings with increased statistics in the future. On the other hand, the $p$-values obtained in the studies involving neutrino tracks are large, being totally compatible with the absence of correlations between the arrival directions of the neutrino tracks  and the UHECRs. Similarly, the $p$-values obtained from the stacking analyses with the point-source sample and the 27 highest-energy UHECRs show no significant excess of neutrinos in the directions of these UHECRs. 

We note that the absence of strong correlations between neutrinos and UHECRs may have different causes. In particular, one has to keep in mind that if the strong suppression in the cosmic-ray flux observed above 40--50~EeV is due to the attenuation of protons in the CMB through photopion production (or of nuclei due to photodisintegrations), the majority of the observed UHECRs above 50~EeV should have been produced in sources relatively nearby, i.e., within $\sim 200$~Mpc. On the other hand, the astrophysical neutrinos may come from sources at any distance since neutrinos are not limited by absorption. Hence, only a small fraction of the extragalactic neutrinos, those from sources relatively nearby, may have the same sources as the observed UHECRs. Considering for instance that the neutrino sources extend to at least z=1, the overall extragalactic contribution arising from sources closer than 200~Mpc is not expected to exceed a few percent (in the case of no source evolution and uniform source distribution) and could be particularly suppressed if the evolution of the sources is strong. Another difficulty in observing common UHECR and neutrino sources is that while the neutrinos arrive straight from their sources, the deflections of charged cosmic rays in Galactic and extragalactic magnetic fields induce sizeable delays between the arrival times of CRs and those of the neutrinos, which are typically much larger than the duration of the experiments. Hence, if the sources are burst-like rather than steady, the burst of neutrinos would arrive long before the associated UHECRs and their simultaneous observation could not be feasible. Finally, the sources capable of producing PeV neutrinos, which are those able to accelerate protons up to energies of a few tens of PeV, may only rarely  be able to also accelerate cosmic rays to ultrahigh-energies. Hence, it may be natural to expect that only a small fraction of the observed energetic neutrinos come from sources that also produce UHECRs. 

In conclusion, further insight will potentially arise from increased statistics and eventually with the inclusion of CR composition information that may become available so as to better model possible effects of magnetic field deflections. This may help to understand if there is a contribution in the astrophysical neutrino signal observed by IceCube correlated to the sources of the observed UHECR.

\acknowledgments
The IceCube Collaboration acknowledges the support from the following agencies:
U.S. National Science Foundation-Office of Polar Programs,
U.S. National Science Foundation-Physics Division,
University of Wisconsin Alumni Research Foundation,
the Grid Laboratory Of Wisconsin (GLOW) grid infrastructure at the University of Wisconsin - Madison, the Open Science Grid (OSG) grid infrastructure;
U.S. Department of Energy, and National Energy Research Scientific Computing Center,
the Louisiana Optical Network Initiative (LONI) grid computing resources;
Natural Sciences and Engineering Research Council of Canada,
WestGrid and Compute/Calcul Canada;
Swedish Research Council,
Swedish Polar Research Secretariat,
Swedish National Infrastructure for Computing (SNIC),
and Knut and Alice Wallenberg Foundation, Sweden;
German Ministry for Education and Research (BMBF),
Deutsche Forschungsgemeinschaft (DFG),
Helmholtz Alliance for Astroparticle Physics (HAP),
Research Department of Plasmas with Complex Interactions (Bochum), Germany;
Fund for Scientific Research (FNRS-FWO),
FWO Odysseus programme,
Flanders Institute to encourage scientific and technological research in industry (IWT),
Belgian Federal Science Policy Office (Belspo);
University of Oxford, United Kingdom;
Marsden Fund, New Zealand;
Australian Research Council;
Japan Society for Promotion of Science (JSPS);
the Swiss National Science Foundation (SNSF), Switzerland;
National Research Foundation of Korea (NRF);
Danish National Research Foundation, Denmark (DNRF).

\noindent The successful installation, commissioning, and operation of the Pierre Auger
Observatory would not have been possible without the strong commitment and
effort from the technical and administrative staff in Malarg\"ue. We are
very grateful to the following agencies and organizations for financial
support:

\begin{sloppypar}
\noindent Comisi\'on Nacional de Energ\'{\i}a At\'omica,
Agencia Nacional de Promoci\'on Cient\'{\i}fica y Tecnol\'ogica (ANPCyT),
Consejo Nacional de Investigaciones Cient\'{\i}ficas y T\'ecnicas (CONICET),
Gobierno de la Provincia de Mendoza,
Municipalidad de Malarg\"ue,
NDM Holdings and Valle Las Le\~nas, in gratitude for their continuing cooperation over land access,
Argentina;
the Australian Research Council;
Conselho Nacional de Desenvolvimento Cient\'{\i}fico e Tecnol\'ogico (CNPq), Financiadora de Estudos e Projetos (FINEP),
Funda\c{c}\~ao de Amparo \`a Pesquisa do Estado de Rio de Janeiro (FAPERJ),
S\~ao Paulo Research Foundation (FAPESP) Grants No.\ 2010/07359-6 and No.\ 1999/05404-3,
Minist\'erio de Ci\^encia e Tecnologia (MCT),
Brazil;
Grant No.\ MSMT-CR LG13007, No.\ 7AMB14AR005, and the Czech Science Foundation Grant No.\ 14-17501S,
Czech Republic;
Centre de Calcul IN2P3/CNRS, Centre National de la Recherche Scientifique (CNRS),
Conseil R\'egional Ile-de-France,
D\'epartement Physique Nucl\'eaire et Corpusculaire (PNC-IN2P3/CNRS),
D\'epartement Sciences de l'Univers (SDU-INSU/CNRS),
Institut Lagrange de Paris (ILP) Grant No.\ LABEX ANR-10-LABX-63,
within the Investissements d'Avenir Programme Grant No.\ ANR-11-IDEX-0004-02,
France;
Bundesministerium f\"ur Bildung und Forschung (BMBF),
Deutsche Forschungsgemeinschaft (DFG),
Finanzministerium Baden-W\"urttemberg,
Helmholtz Alliance for Astroparticle Physics (HAP),
Helmholtz-Gemeinschaft Deutscher Forschungszentren (HGF),
Ministerium f\"ur Wissenschaft und Forschung, Nordrhein Westfalen,
Ministerium f\"ur Wissenschaft, Forschung und Kunst, Baden-W\"urttemberg,
Germany;
Istituto Nazionale di Fisica Nucleare (INFN),
Istituto Nazionale di Astrofisica (INAF),
Ministero dell'Istruzione, dell'Universit\'a e della Ricerca (MIUR),
Gran Sasso Center for Astroparticle Physics (CFA),
CETEMPS Center of Excellence, Ministero degli Affari Esteri (MAE),
Italy;
Consejo Nacional de Ciencia y Tecnolog\'{\i}a (CONACYT),
Mexico;
Ministerie van Onderwijs, Cultuur en Wetenschap,
Nederlandse Organisatie voor Wetenschappelijk Onderzoek (NWO),
Stichting voor Fundamenteel Onderzoek der Materie (FOM),
Netherlands;
National Centre for Research and Development, Grants No.\ ERA-NET-ASPERA/01/11 and No.\ ERA-NET-ASPERA/02/11,
National Science Centre, Grants No.\ 2013/08/M/ST9/00322, No.\ 2013/08/M/ST9/00728 and No.\ HARMONIA 5 - 2013/10/M/ST9/00062,
Poland;
Portuguese national funds and FEDER funds within Programa Operacional Factores de Competitividade through Funda\c{c}\~ao para a Ci\^encia e a Tecnologia (COMPETE),
Portugal;
Romanian Authority for Scientific Research ANCS,
CNDI-UEFISCDI partnership projects Grants No.\ 20/2012 and No.\ 194/2012,
Grants No.\ 1/ASPERA2/2012 ERA-NET, No.\ PN-II-RU-PD-2011-3-0145-17 and No.\ PN-II-RU-PD-2011-3-0062,
the Minister of National Education,
Programme Space Technology and Advanced Research (STAR), Grant No.\ 83/2013,
Romania;
Slovenian Research Agency,
Slovenia;
Comunidad de Madrid,
FEDER funds,
Ministerio de Educaci\'on y Ciencia,
Xunta de Galicia,
European Community 7th Framework Program, Grant No.\ FP7-PEOPLE-2012-IEF-328826,
Spain;
Science and Technology Facilities Council,
United Kingdom;
Department of Energy, Contracts No.\ DE-AC02-07CH11359, No.\ DE-FR02-04ER41300, No.\ DE-FG02-99ER41107 and No.\ DE-SC0011689,
National Science Foundation, Grant No.\ 0450696,
The Grainger Foundation,
USA;
NAFOSTED,
Vietnam;
Marie Curie-IRSES/EPLANET,
European Particle Physics Latin American Network,
European Union 7th Framework Program, Grant No.\ PIRSES-2009-GA-246806 and PIOF-GA-2013-624803;
and
UNESCO.
\end{sloppypar}

\begin{sloppypar}
\noindent The Telescope Array experiment is supported by the Japan Society for the
Promotion of Science through Grants-in-Aid for Scientific Research on Specially
Promoted Research (21000002) ``Extreme Phenomena in the Universe Explored by
Highest Energy Cosmic Rays'' and for Scientific Research (19104006), and the
Inter-University Research Program of the Institute for Cosmic Ray Research;
by the U.S.\ National Science Foundation awards PHY-0307098, PHY-0601915,
PHY-0649681, PHY-0703893, PHY-0758342, PHY-0848320, PHY-1069280, PHY-1069286,
PHY-1404495 and PHY-1404502;
by the National Research Foundation of Korea (2007-0093860, R32-10130,
2012R1A1A2008381, 2013004883);
by
the Russian Academy of Sciences, RFBR grants 11-02-01528a and 13-02-01311a
(INR), IISN project No.\ 4.4502.13, and Belgian Science Policy under IUAP VII/37
(ULB).
The foundations of Dr.~Ezekiel R.~and Edna Wattis Dumke, Willard L.\ Eccles,
and George S.\ and Dolores Dor\'e Eccles all helped with generous donations.
The State of Utah supported the project through its Economic Development Board,
and the University of Utah through the Office of the Vice President for
Research.
The experimental site became available through the cooperation of the Utah
School and Institutional Trust Lands Administration (SITLA), U.S.\ Bureau of
Land Management, and the U.S.\ Air Force. We also wish to thank the people and
the officials of Millard County, Utah for their steadfast and warm support.
We gratefully acknowledge the contributions from the technical staffs of our
home institutions.
An allocation of computer time from the Center for High Performance Computing
at the University of Utah is gratefully acknowledged.
\end{sloppypar}

\bibliographystyle{model1a-num-names}
\bibliography{<your-bib-database>}

\begin{thebibliography}{100}

\bibitem{augerphotons} J.~Abraham {\it et al.} [The Pierre Auger Collaboration], Phys. Rev. D {\bf 79}  (2009) 102001.

\bibitem{taphotons}  T.~Abu-Zayyad {\it et al.} [Telescope Array Collaboration], Phys. Rev. D {\bf 88}  (2013) 112005.

\bibitem{augernim04} J.~Abraham {\it et al.} [The Pierre Auger Collaboration], Nucl. Instrum. Meth. A {\bf523} (2004) 50.

\bibitem{augernim15} A.~Aab {\it et al.} [The Pierre Auger Collaboration], Nucl. Instrum. Meth. A {\bf 798} (2015) 172.

\bibitem{AbuZayyad:2012kk} T.~Abu-Zayyad {\it et al.}  [Telescope Array Collaboration], Nucl. Instrum. Meth. A {\bf 689} (2012) 87.

\bibitem{augerspectrum} J.~Abraham {\it et al.}  [The Pierre Auger Collaboration], Phys. Rev. Lett. {\bf 101} (2008) 061101.

\bibitem{taspectrum} T.~Abu-Zayyad {\it et al.} [The Telescope Array Collaboration], Astrophys. J. {\bf 768} (2013) L1.

\bibitem{gzk} K.~Greisen, Phys. Rev. Lett {\bf 16} (1966) 748; G.T.~Zatsepin and V.A.~Kuz'min, J. Exp. Theor. Phys. Lett. (1966) 78. 

\bibitem{augerapj2015} A.~Aab {\it et al.}, [The Pierre Auger Collaboration], Astrophys. J. {\bf 804} (2015) 15.

\bibitem{taardir} T.~Abu-Zayyad {\it et al.} [The Telescope Array Collaboration], Astrophys. J. {\bf 777} (2013) 88.

\bibitem{farrarcc2014} G.~Farrar, Comptes Rendus Physique {\bf 15} (2014) 339.

\bibitem{Haverkorn:2014jka} M.~Haverkorn, {\it Magnetic Fields in the Milky Way}, to appear in, ``Magnetic Fields in Diffuse Media", eds. E.M.~de Gouveia Dal Pino and A.~Lazarian, arXiv:1406.0283.

\bibitem{augerxmax} A.~Aab {\it et al.} [The Pierre Auger Collaboration], Phys. Rev. D {\bf 90} (2014) 12 122005.

\bibitem{taxmax} R.U.~Abbasi {\it et al.} [The Telescope Array Collaboration], Astropart. Phys. {\bf 64} (2014) 49.

\bibitem{HESE2paper} M.G.~Aartsen {\it et al.} [The IceCube Collaboration], Science {\bf 342} (2013) 1242856. 

\bibitem{olinto} K.~Fang, T.~Fujii, T.~Linden and A.V.~Olinto, Astrophys. J. {\bf 794} (2014) 126.

\bibitem{razzaque} R.~Moharana, S.~Razzaque, J. Cosmology \& Astropart. Phys.  {\bf 08} (2015) 014.

\bibitem{gago} J.A.~Carpio, A.M.~Gago, arXiv:1507.02781.

\bibitem{waxbah} E.~Waxman and J.N.~Bahcall, Phys. Rev. D {\bf 59} (1999) 023002.

\bibitem{mannheim} K.~Mannheim, R.J.~Protheroe and J.P.~Rachen, Phys. Rev. D {\bf 63} (2001) 023003.

\bibitem{HESE3paper} M.G.~Aartsen {\it et al.} [The IceCube Collaboration], Phys. Rev. Lett. {\bf 113} (2014) 101101.

\bibitem{HESE4yrICRC} M.G.~Aartsen {\it et al.} [The IceCube Collaboration], Proc. of the $34^{th}$ Int. Cosmic Ray Conf. 1081 (2015), arXiv:1510.05223.

\bibitem{weaver} M.G.~Aartsen {\it et al.} [The IceCube Collaboration],  Phys. Rev. Lett. {\bf 115} (2015) 8 081102.

\bibitem{IC86PSPaper} M.G.~Aartsen {\it et al.} [The IceCube Collaboration],  Astrophys. J. {\bf 796} (2014) 109.

\bibitem{Utah} A.~Christov, G.~Golup, T.~Montaruli, M.~Rameez for the IceCube Collaboration, J.~Aublin, L.~Caccianiga, P.L.~Ghia, E.~Roulet, M.~Unger for the Pierre Auger Collaboration, and H.~Sagawa, P.~Tinyakov for the Telescope Array Collaboration,  {\it Towards a Joint Analysis of Data from the IceCube Neutrino Telescope, the Pierre Auger Observatory and Telescope Array}, Proc. of the UHECR 2014 Symposium, to appear in JPS Conf. Proc.

\bibitem[Achterberg {\it et al.}(2006)]{FirstYearPerformancePaper} A.~Achterberg {\it et al.} [The IceCube Collaboration], Astropart. Phys. {\bf 26} (2006) 155.

\bibitem[Abbasi {\it et al.}(2010)]{PMTPaper} R.~Abbasi {\it et al.} [The IceCube Collaboration], Nucl. Instrum. Meth. A {\bf 618} (2010) 139.

\bibitem[Abbasi {\it et al.}(2009)]{DOMMBPaper} R.~Abbasi {\it et al.} [The IceCube Collaboration], Nucl. Instrum. Meth. A {\bf 601} (2009) 294.

\bibitem{CWdatarelease} https://icecube.wisc.edu/science/data/HE$\_$NuMu$\_$diffuse

\bibitem[Aartsen {\it et al.}(2013c)]{IC79Paper} M.G.~Aartsen {\it et al.} [The IceCube Collaboration],  Astrophys. J. {\bf 779} (2013) 132.  

\bibitem[Abbasi {\it et al.}(2011)]{IC40Paper} R.~Abbasi {\it et al.} [The IceCube Collaboration], Astrophys. J. {\bf 732} (2011) 18.

\bibitem{augerar} C.~Bonifazi {\it et al.} [The Pierre Auger Collaboration], Nucl. Phys. B Proc. Suppl.  {\bf 190} (2009) 20.

\bibitem{augerrec1} M.~Ave {\it et al.} [The Pierre Auger Collaboration], Proc. of the $30^{th}$ Int. Cosmic Ray Conf. (2007) {\bf 4} 307, arXiv:0709.2125.

\bibitem{augerrec2} A.~Aab {\it et al.} [The Pierre Auger Collaboration], J. Cosmology \& Astropart. Phys.  {\bf 8} (2014) 19.

\bibitem{augerene1} V.~Verzi {\it et al.} [The Pierre Auger Collaboration], Proc. of the $33^{rd}$ Int. Cosmic Ray Conf. (2013), arXiv:1307.5059.

\bibitem{augerene2} R.~Pesce {\it et al.} [The Pierre Auger Collaboration], Proc. of the $32^{nd}$ Int. Cosmic Ray Conf. (2011), arXiv:1107.4809.

\bibitem{augernim10} J.~Abraham {\it et al.} [The Pierre Auger Collaboration], Nucl. Instrum. Meth. A {\bf 613} (2010) 29.

\bibitem{tahotspot} R.U.~Abbasi {\it et al.}  [The Telescope Array Collaboration], Astrophys. J. {\bf 790} (2014) L21.

\bibitem{UHECRMaris} I.~Maris for the Telescope Array and Pierre Auger Collaborations, {\it High Energy Spectrum Working Group Report: The energy spectrum of ultra high energy cosmic rays},  Proc. of the UHECR 2014 Symposium, to appear in JPS Conf. Proc.

\bibitem{Kronberg:1993vk} P.P.~Kronberg,  Rept. Prog. Phys. {\bf 57} (1994) 325.

\bibitem{Durrer:2013pga} R.~Durrer and A.~Neronov, Astron. Astrophys. Rev.  {\bf 21} (2013) 62.

\bibitem{Pshirkov:2015tua} M.S.~Pshirkov, P.G.~Tinyakov and F.R.~Urban, arXiv:1504.06546.

\bibitem{Pshirkov:2011um} M.S.~Pshirkov, P.G.~Tinyakov, P.P.~Kronberg and K.J.~Newton-McGee,  Astrophys. J. {\bf 738}  (2011) 192.

\bibitem{Jansson:2012pc} R.~Jansson and G.R.~Farrar,  Astrophys. J. {\bf 757} (2012) 14.

\bibitem{random-deflections} P.G.~Tinyakov and I.I.~Tkachev, Astropart. Phys. {\bf 24} (2005) 32.

\bibitem{Augerindex} M.~Settimo {\it et al.} [The Pierre Auger Collaboration], Eur. Phys. J. Plus {\bf 87} (2012) 127.

\bibitem{TAindex} D.~Ivanov for the Telescope Array Collaboration, {\it High Energy Spectrum Measured by the Telescope Array Experiment}, Proc. of the UHECR 2014 Symposium, to appear in JPS Conf. Proc.

\bibitem{Wilkstheorem} S.S.~Wilks,  Ann. Math. Statist. {\bf 9} (1938) 1.

\bibitem{method} J.~Braun {\it et al.}, Astropart. Phys. {\bf29} (2008) 299.

\end{thebibliography}

\newpage
\appendix
\section{Neutrino }
\begin{table}[!h]
\caption{List of the neutrino cascade-like events used in the analyses. The ID number corresponds to the ID in \cite{HESE3paper,HESE4yrICRC}. `Dep. Energy' corresponds to the EM equivalent energy deposited within the detector.}
\label{cascades_list}
\begin{center}
\begin{tabular}{*5c}
\hline
ID & Dep. Energy [TeV]             & RA [$^\circ$]          &  dec [$^\circ$]& Med. Angular Error [$^\circ$]  \\
 \hline
 1   & $47.6^{+6.5}_{-5.4}$        & 35.2			&-1.8              & 16.3 \\
 2   & $117^{+15}_{-15}$           & 282.6			&-28.0            & 25.4 \\
 4   & $165^{+20}_{-15}$           & 169.5			&-51.2            & 7.1 \\
 6   & $28.4^{+2.7}_{-2.5}$        &133.9			&-27.2            & 9.8 \\
 7   & $34.3^{+3.5}_{-4.3}$        &15.6			&-45.1            & 24.1 \\
 9   & $63.2^{+7.1}_{-8.0}$        &151.3			&33.6             & 16.5 \\
10 & $97.2^{+10.4}_{-12.4}$     &5.0				&-29.4            & 8.1 \\
11 & $88.4^{+12.5}_{-10.7}$     &155.3			&-8.9              & 16.7 \\
12 & $104^{+13}_{-13}$            &296.1			&-52.8            & 9.8 \\
14 & $1041^ {+132}_{-144}$     & 265.6			&-27.9            &13.2 \\
15 & $57.5^{+8.3}_{-7.8}$         & 287.3			&-49.7            & 19.7 \\
16 & $30.6^{+3.6}_{-3.5}$         & 192.1			&-22.6            & 19.4 \\
17 & $200^{+27}_{-27}$            & 247.4			&14.5             & 11.6 \\
19 & $71.5^{+7.0}_{-7.2}$         & 76.9			&-59.7            & 9.7 \\
20 & $1141^{+143}_{-133}$      & 38.3			&-67.2            &10.7 \\
21 & $30.2^{+3.5}_{-3.3}$         & 9.0				&-24.0            & 20.9 \\
22 & $220^{+21}_{-24}$            & 293.7			&-22.1            & 12.1 \\
24 & $30.5^{+3.2}_{-2.6}$         & 282.2			&-15.1            & 15.5 \\
25 & $33.5^{+4.9}_{-5.0}$         & 286.0			&-14.5            & 46.3 \\
26 & $210^{+29}_{-26}$            & 143.4			& 22.7            & 11.8 \\
27 & $60.2^{+5.6}_{-5.6}$         &121.7			&-12.6            & 6.6 \\
29 & $32.7^{+3.2}_{-2.9}$         & 298.1			& 41.0            & 7.4 \\
30 & $129^{+14}_{-12}$            &103.2			&-82.7            & 8.0 \\
31 & $42.5 ^{+5.4}_{-5.7}$        &146.1			& 78.3            & 26.0 \\
33 & $385^{+46}_{-49}$            & 292.5			& 7.8              & 13.5 \\
34 & $42.1^{+6.5}_{-6.3}$         	& 323.4			& 31.3            & 42.7 \\
35 & $2004^{+236}_{-262}$      & 208.4			&-55.8            & 15.9 \\
36 & $28.9^{+3.0}_{-2.6}$         & 257.7			&-3.0              & 11.7 \\
39 & $101.3^{+13.3}_{-11.6}$   &106.2                       &-17.9            &14.2\\
40 & $157.3^{+15.9}_{-16.7}$   &143.9                       &-48.5            &11.7\\
41 & $87.6^{+8.4}_{-10.0}$       &66.1                         &3.3               &11.1\\
42 & $76.3^{+10.3}_{-11.6}$     &42.5                         &-25.3            &20.7\\
46 & $158.0^{+15.3}_{-16.6}$   &150.5                       &-22.4            &7.6\\
48 & $104.7^{+13.5}_{-10.2}$   &213.1                       &-33.2            &8.1\\
49 & $59.9^{+8.3}_{-7.9}$         &203.2                       &-26.3            &21.8\\
50 & $22.2^{+2.3}_{-2.0}$         &168.6                       &59.3             &8.2\\
51 & $66.2^{+6.7}_{-6.1}$         &88.6                         &54.0             &6.5\\
52 & $158.1^{+16.3}_{18.4}$    &252.8                       &-54.0            &7.8\\
54 & $54.5^{+5.1}_{-6.3}$         &170.5                       &6.0               &11.6\\
 \hline
 \end{tabular}
 \end{center}
\end{table}

\begin{table}[!h]
\caption{List of the neutrino track-like events used in the analyses. The first seven ID numbers correspond to the IDs in \cite{HESE3paper,HESE4yrICRC}. The rest of the ID numbers correspond to the high-energy tracks from \cite{weaver} described in Section 2.1.2. For these events, the `most probable muon energy' is an estimate of the muon energy assuming the best fit spectrum resulting from the flux fit in that analysis \cite{weaver}, which is different from the deposited energy specified for HESE tracks and cascades.}
\label{tracks_list}
\begin{center}
\begin{tabular}{*5c}
\hline
ID & Dep. Energy [TeV]            & RA [$^\circ$]          &  dec [$^\circ$] & Med. Angular Error [$^\circ$]  \\
\hline
 5   & $71.4^{+9.0}_{-9.0}$	 &110.6			&-0.4		&$\lesssim1.2$\\
13  &$253^{+26}_{-22}$		 &67.9			&40.3		&$\lesssim1.2$\\
 23 &$82.2^{+8.6}_{-8.4}$		 &208.7			&-13.2		&$\lesssim1.9$\\
 38 &$200.5^{+16.4}_{-16.4}$    &93.3                        &14.0               &$\lesssim1.2$\\
 44 &$84.6^{+7.4}_{-7.9}$          &336.7                      &0.0                 &$\lesssim1.2$\\
 45 &$429.9^{+57.4}_{-49.1}$    &219.0                      &-86.3              &$\lesssim1.2$\\
 47 &$74.3^{+8.3}_{-7.2}$          &209.4                      &67.4               &$\lesssim1.2$\\
 \hline
ID & Most probable muon energy [TeV]            & RA [$^\circ$]          &  dec [$^\circ$] & Med. Angular Error [$^\circ$]  \\
\hline		
116357,6324295     &	 755		&254	.0   		&16.3	&$\lesssim1.0$\\       	    	  
116807,9493609     &	 604		&88.5	   	&0.2		&$\lesssim1.0$\\		           
119136,66932419   &	397		&37.1	  	&18.6	&$\lesssim1.0$\\	  
116883,17395151   &	422 		&285.7	 	&3.1		&$\lesssim1.0$\\	   
116701,6581938     &	 317		&331	.0    		&11.0    	&$\lesssim1.0$\\
116026,44241207   &	339		&346.8	 	&24.0	&$\lesssim1.0$\\	    
116574,20123342   &	302		&267.5	 	&13.8	&$\lesssim1.0$\\	  
119739,41603205   &	326		&238.3  		&18.9	&$\lesssim1.0$\\	  
118210,47538807   &	252		&235.2	 	&19.3	&$\lesssim1.0$\\	  
\hline
\end{tabular}
\end{center}
\end{table}

\newpage
\noindent{\bf The IceCube Collaboration}
\begin{sloppypar}
\noindent M.~G.~Aartsen$^{2}$,
K.~Abraham$^{32}$,
M.~Ackermann$^{48}$,
J.~Adams$^{15}$,
J.~A.~Aguilar$^{12}$,
M.~Ahlers$^{29}$,
M.~Ahrens$^{39}$,
D.~Altmann$^{23}$,
T.~Anderson$^{45}$,
I.~Ansseau$^{12}$
% ...
\\
% ...
$^{2}$ Department of Physics, University of Adelaide, Adelaide, 5005, Australia \\
% ...
$^{12}$ Universit\'e Libre de Bruxelles, Science Faculty CP230, B-1050 Brussels, Belgium \\
% ...
$^{15}$ Dept.~of Physics and Astronomy, University of Canterbury, Private Bag 4800, Christchurch, New Zealand \\
% ...
$^{23}$ Erlangen Centre for Astroparticle Physics, Friedrich-Alexander-Universit\"at Erlangen-N\"urnberg, D-91058 Erlangen, Germany \\
% ...
$^{29}$ Dept.~of Physics and Wisconsin IceCube Particle Astrophysics Center, University of Wisconsin, Madison, WI 53706, USA \\
% ...
$^{32}$ Technische Universit\"at M\"unchen, D-85748 Garching, Germany \\
% ...
$^{39}$ Oskar Klein Centre and Dept.~of Physics, Stockholm University, SE-10691 Stockholm, Sweden \\
% ...
$^{45}$ Dept.~of Physics, Pennsylvania State University, University Park, PA 16802, USA \\
% ...
$^{48}$ DESY, D-15735 Zeuthen, Germany \\
% ...
\end{sloppypar}
\newpage
\noindent{\bf The Pierre Auger Collaboration}
\begin{sloppypar}
\noindent A.~Aab$^{40}$,
P.~Abreu$^{71}$,
M.~Aglietta$^{50,49}$,
E.J.~Ahn$^{86}$,
I.~Al Samarai$^{30}$,
I.F.M.~Albuquerque$^{16}$,
I.~Allekotte$^{1}$,
P.~Allison$^{91}$,
A.~Almela$^{11,8}$,
J.~Alvarez Castillo$^{64}$
% ...
\\
\llap{$^{1}$} Centro At\'omico Bariloche and Instituto Balseiro (CNEA-UNCuyo-CONICET), Argentina\\
\llap{$^{2}$} Centro de Investigaciones en L\'aseres y Aplicaciones, CITEDEF and CONICET, Argentina\\
\llap{$^{3}$} Departamento de F\'\i{}sica, FCEyN, Universidad de Buenos Aires, Argentina\\
\llap{$^{4}$} IFLP, Universidad Nacional de La Plata and CONICET, Argentina\\
\llap{$^{5}$} Instituto de Astronom\'\i{}a y F\'\i{}sica del Espacio (IAFE, CONICET-UBA), Argentina\\
\llap{$^{6}$} Instituto de F\'\i{}sica de Rosario (IFIR) -- CONICET/U.N.R.\ and Facultad de Ciencias Bioqu\'\i{}micas y Farmac\'euticas U.N.R., Argentina\\
\llap{$^{7}$} Instituto de Tecnolog\'\i{}as en Detecci\'on y Astropart\'\i{}culas (CNEA, CONICET, UNSAM) and Universidad Tecnol\'ogica Nacional -- Facultad Regional Mendoza (CONICET/CNEA), Argentina\\
\llap{$^{8}$} Instituto de Tecnolog\'\i{}as en Detecci\'on y Astropart\'\i{}culas (CNEA, CONICET, UNSAM), Centro At\'omico Constituyentes, Comisi\'on Nacional de Energ\'\i{}a At\'omica, Argentina\\
\llap{$^{9}$} Observatorio Pierre Auger, Argentina\\
\llap{$^{10}$} Observatorio Pierre Auger and Comisi\'on Nacional de Energ\'\i{}a At\'omica, Argentina\\
\llap{$^{11}$} Universidad Tecnol\'ogica Nacional -- Facultad Regional Buenos Aires, Argentina\\
\llap{$^{12}$} University of Adelaide, Australia\\
\llap{$^{13}$} Centro Brasileiro de Pesquisas Fisicas (CBPF), Brazil\\
\llap{$^{14}$} Universidade de S\~ao Paulo, Escola de Engenharia de Lorena, Brazil\\
\llap{$^{15}$} Universidade de S\~ao Paulo, Inst.\ de F\'\i{}sica de S\~ao Carlos, S\~ao Carlos, Brazil\\
\llap{$^{16}$} Universidade de S\~ao Paulo, Inst.\ de F\'\i{}sica, S\~ao Paulo, Brazil\\
\llap{$^{17}$} Universidade Estadual de Campinas (UNICAMP), Brazil\\
\llap{$^{18}$} Universidade Estadual de Feira de Santana (UEFS), Brazil\\
\llap{$^{19}$} Universidade Federal da Bahia, Brazil\\
\llap{$^{20}$} Universidade Federal de Pelotas, Brazil\\
\llap{$^{21}$} Universidade Federal do ABC (UFABC), Brazil\\
\llap{$^{22}$} Universidade Federal do Rio de Janeiro (UFRJ), Instituto de F\'\i{}sica, Brazil\\
\llap{$^{23}$} Universidade Federal Fluminense, Brazil\\
\llap{$^{24}$} Universidad Industrial de Santander, Colombia\\
\llap{$^{25}$} Institute of Physics (FZU) of the Academy of Sciences of the Czech Republic, Czech Republic\\
\llap{$^{26}$} Palacky University, RCPTM, Czech Republic\\
\llap{$^{27}$} University Prague, Institute of Particle and Nuclear Physics, Czech Republic\\
\llap{$^{28}$} Institut de Physique Nucl\'eaire d'Orsay (IPNO), Universit\'e Paris 11, CNRS-IN2P3, France\\
\llap{$^{29}$} Laboratoire de l'Acc\'el\'erateur Lin\'eaire (LAL), Universit\'e Paris 11, CNRS-IN2P3, France\\
\llap{$^{30}$} Laboratoire de Physique Nucl\'eaire et de Hautes Energies (LPNHE), Universit\'es Paris 6 et Paris 7, CNRS-IN2P3, France\\
\llap{$^{31}$} Laboratoire de Physique Subatomique et de Cosmologie (LPSC), Universit\'e Grenoble-Alpes, CNRS/IN2P3, France\\
\llap{$^{32}$} Station de Radioastronomie de Nan\c{c}ay, France\\
\llap{$^{33}$} SUBATECH, \'Ecole des Mines de Nantes, CNRS-IN2P3, Universit\'e de Nantes, France\\
\llap{$^{34}$} Bergische Universit\"at Wuppertal, Fachbereich C -- Physik, Germany\\
\llap{$^{35}$} Karlsruhe Institute of Technology, Institut f\"ur Experimentelle Kernphysik (IEKP), Germany\\
\llap{$^{36}$} Karlsruhe Institute of Technology, Institut f\"ur Kernphysik (IKP), Germany\\
\llap{$^{37}$} Karlsruhe Institute of Technology, Institut f\"ur Prozessdatenverarbeitung und Elektronik (IPE), Germany\\
\llap{$^{38}$} RWTH Aachen University, III.\ Physikalisches Institut A, Germany\\
\llap{$^{39}$} Universit\"at Hamburg, II.\ Institut f\"ur Theoretische Physik, Germany\\
\llap{$^{40}$} Universit\"at Siegen, Fachbereich 7 Physik -- Experimentelle Teilchenphysik, Germany\\
\llap{$^{41}$} INAF -- Istituto di Astrofisica Spaziale e Fisica Cosmica di Palermo, Italy\\
\llap{$^{42}$} INFN Laboratori del Gran Sasso, Italy\\
\llap{$^{43}$} INFN, Sezione di Catania, Italy\\
\llap{$^{44}$} INFN, Sezione di L'Aquila, Italy\\
\llap{$^{45}$} INFN, Sezione di Lecce, Italy\\
\llap{$^{46}$} INFN, Sezione di Milano, Italy\\
\llap{$^{47}$} INFN, Sezione di Napoli, Italy\\
\llap{$^{48}$} INFN, Sezione di Roma "Tor Vergata", Italy\\
\llap{$^{49}$} INFN, Sezione di Torino, Italy\\
\llap{$^{50}$} Osservatorio Astrofisico di Torino (INAF), Torino, Italy\\
\llap{$^{51}$} Universit\`a Torino, Dipartimento di Fisica, Italy\\
\llap{$^{52}$} Universit\`a del Salento, Dipartimento di Ingegneria, Italy\\
\llap{$^{53}$} Universit\`a del Salento, Dipartimento di Matematica e Fisica ``E.\ De Giorgi'', Italy\\
\llap{$^{54}$} Universit\`a dell'Aquila, Dipartimento di Chimica e Fisica, Italy\\
\llap{$^{55}$} Universit\`a di Catania, Dipartimento di Fisica e Astronomia, Italy\\
\llap{$^{56}$} Universit\`a di Milano, Dipartimento di Fisica, Italy\\
\llap{$^{57}$} Universit\`a di Napoli "Federico II", Dipartimento di Fisica, Italy\\
\llap{$^{58}$} Universit\`a di Roma ``Tor Vergata'', Dipartimento di Fisica, Italy\\
\llap{$^{59}$} Benem\'erita Universidad Aut\'onoma de Puebla (BUAP), M\'exico\\
\llap{$^{60}$} Centro de Investigaci\'on y de Estudios Avanzados del IPN (CINVESTAV), M\'exico\\
\llap{$^{61}$} Unidad Profesional Interdisciplinaria en Ingenier\'\i{}a y Tecnolog\'\i{}as Avanzadas del Instituto Polit\'ecnico Nacional (UPIITA-IPN), M\'exico\\
\llap{$^{62}$} Universidad Aut\'onoma de Chiapas, M\'exico\\
\llap{$^{63}$} Universidad Michoacana de San Nicol\'as de Hidalgo, M\'exico\\
\llap{$^{64}$} Universidad Nacional Aut\'onoma de M\'exico, M\'exico\\
\llap{$^{65}$} Institute for Mathematics, Astrophysics and Particle Physics (IMAPP), Radboud Universiteit, Nijmegen, Netherlands\\
\llap{$^{66}$} KVI -- Center for Advanced Radiation Technology, University of Groningen, Netherlands\\
\llap{$^{67}$} Nationaal Instituut voor Kernfysica en Hoge Energie Fysica (NIKHEF), Netherlands\\
\llap{$^{68}$} Stichting Astronomisch Onderzoek in Nederland (ASTRON), Dwingeloo, Netherlands\\
\llap{$^{69}$} Institute of Nuclear Physics PAN, Poland\\
\llap{$^{70}$} University of \L{}\'od\'z, Poland\\
\llap{$^{71}$} Laborat\'orio de Instrumenta\c{c}\~ao e F\'\i{}sica Experimental de Part\'\i{}culas -- LIP and Instituto Superior T\'ecnico -- IST, Universidade de Lisboa -- UL, Portugal\\
\llap{$^{72}$} ``Horia Hulubei'' National Institute for Physics and Nuclear Engineering, Romania\\
\llap{$^{73}$} Institute of Space Science, Romania\\
\llap{$^{74}$} University of Bucharest, Physics Department, Romania\\
\llap{$^{75}$} University Politehnica of Bucharest, Romania\\
\llap{$^{76}$} Experimental Particle Physics Department, J.\ Stefan Institute, Slovenia\\
\llap{$^{77}$} Laboratory for Astroparticle Physics, University of Nova Gorica, Slovenia\\
\llap{$^{78}$} Universidad Complutense de Madrid, Spain\\
\llap{$^{79}$} Universidad de Alcal\'a de Henares, Spain\\
\llap{$^{80}$} Universidad de Granada and C.A.F.P.E., Spain\\
\llap{$^{81}$} Universidad de Santiago de Compostela, Spain\\
\llap{$^{82}$} Case Western Reserve University, USA\\
\llap{$^{83}$} Colorado School of Mines, USA\\
\llap{$^{84}$} Colorado State University, USA\\
\llap{$^{85}$} Department of Physics and Astronomy, Lehman College, City University of New York, USA\\
\llap{$^{86}$} Fermi National Accelerator Laboratory, USA\\
\llap{$^{87}$} Louisiana State University, USA\\
\llap{$^{88}$} Michigan Technological University, USA\\
\llap{$^{89}$} New York University, USA\\
\llap{$^{90}$} Northeastern University, USA\\
\llap{$^{91}$} Ohio State University, USA\\
\llap{$^{92}$} Pennsylvania State University, USA\\
\llap{$^{93}$} University of Chicago, USA\\
\llap{$^{94}$} University of Hawaii, USA\\
\llap{$^{95}$} University of Nebraska, USA\\
\llap{$^{96}$} University of New Mexico, USA\\
\llap{$^{a}$} School of Physics and Astronomy, University of Leeds, Leeds, United Kingdom\\
\llap{$^{b}$} Max-Planck-Institut f\"ur Radioastronomie, Bonn, Germany\\
\llap{$^{c}$} also at Vrije Universiteit Brussels, Brussels, Belgium\\
\llap{$^{d}$} also at Universidade Federal do Paran\'a, Palotina, PR, Brazil
\end{sloppypar}

\newpage
\noindent{\bf The Telescope Array Collaboration}
\begin{sloppypar}
\noindent R.U.~Abbasi$^{1}$,
M.~Abe$^{2}$,
T.~Abu-Zayyad$^{1}$,
M.~Allen$^{1}$,
R.~Azuma$^{3}$,
E.~Barcikowski$^{1}$,
J.W.~Belz$^{1}$,
D.R.~Bergman$^{1}$,
S.A.~Blake$^{1}$,
R.~Cady$^{1}$
% ...
\\
$^{1}$ High Energy Astrophysics Institute and Department of Physics and Astronomy, University of Utah, Salt Lake City, Utah, USA\\
$^{2}$ The Graduate School of Science and Engineering, Saitama University, Saitama, Saitama, Japan\\
$^{3}$ Graduate School of Science and Engineering, Tokyo Institute of Technology, Meguro, Tokyo, Japan\\
$^{4}$ Department of Physics and Institute for the Early Universe, Ewha Womans University, Seodaaemun-gu, Seoul, Korea\\
$^{5}$ Department of Physics and The Research Institute of Natural Science, Hanyang University, Seongdong-gu, Seoul, Korea\\
$^{6}$ Department of Physics, Tokyo University of Science, Noda, Chiba, Japan\\
$^{7}$ Department of Physics, Kinki University, Higashi Osaka, Osaka, Japan\\
$^{8}$ Department of Physics, Yonsei University, Seodaemun-gu, Seoul, Korea\\
$^{9}$ Institute for Cosmic Ray Research, University of Tokyo, Kashiwa, Chiba, Japan\\
$^{10}$ Kavli Institute for the Physics and Mathematics of the Universe (WPI), Todai Institutes for Advanced Study, the University of Tokyo, Kashiwa, Chiba, Japan\\
$^{11}$ Graduate School of Science, Osaka City University, Osaka, Osaka, Japan\\
$^{12}$ Faculty of Engineering, Kanagawa University, Yokohama, Kanagawa, Japan\\
$^{13}$ Interdisciplinary Graduate School of Medicine and Engineering, University of Yamanashi, Kofu, Yamanashi, Japan\\
$^{14}$ Astrophysical Big Bang Laboratory, RIKEN, Wako, Saitama, Japan\\
$^{15}$ Department of Physics, Tokyo City University, Setagaya-ku, Tokyo, Japan\\
$^{16}$ Institute for Nuclear Research of the Russian Academy of Sciences, Moscow, Russia\\
$^{17}$ Advanced Research Institute for Science and Engineering, Waseda University, Shinjuku-ku, Tokyo, Japan\\
$^{18}$ Department of Physics, Chiba University, Chiba, Chiba, Japan\\
$^{19}$ Department of Physics, School of Natural Sciences, Ulsan National Institute of Science and Technology, UNIST-gil, Ulsan, Korea\\
$^{20}$ Institute of Particle and Nuclear Studies, KEK, Tsukuba, Ibaraki, Japan\\
$^{21}$ Faculty of Science, Kochi University, Kochi, Kochi, Japan\\
$^{22}$ Department of Physical Sciences, Ritsumeikan University, Kusatsu, Shiga, Japan\\
$^{23}$ Department of Physics, Kyushu University, Fukuoka, Fukuoka, Japan\\
$^{24}$ Engineering Science Laboratory, Chubu University, Kasugai, Aichi, Japan\\
$^{25}$ Department of Physics, Sungkyunkwan University, Jang-an-gu, Suwon, Korea\\
$^{26}$ Sternberg Astronomical Institute,  Moscow M.V.~Lomonosov State University, Moscow, Russia\\
$^{27}$ Department of Physics and Astronomy, Rutgers University -- The State University of New Jersey, Piscataway, New Jersey, USA\\
$^{28}$ Earthquake Research Institute, University of Tokyo, Bunkyo-ku, Tokyo, Japan\\
$^{29}$ Graduate School of Information Sciences, Hiroshima City University, Hiroshima, Hiroshima, Japan\\
$^{30}$ Service de Physique Th\'eorique, Universit\'e Libre de Bruxelles, Brussels, Belgium\\
$^{31}$ Department of Computer Science and Engineering, Shinshu University, Nagano, Nagano, Japan\\
$^{32}$ National Institute of Radiological Science, Chiba, Chiba, Japan\\
$^{33}$ Department of Physics, Ehime University, Matsuyama, Ehime, Japan\\
% ...
\end{sloppypar}

\end{document}